# Survey on Vehicular Ad Hoc Networks and Its Access Technologies Security Vulnerabilities and Countermeasures

Kaveh Bakhsh Kelarestaghi, Mahsa Foruhandeh, Kevin Heaslip, Ryan Gerdes

*Abstract*— In this study, we attempt to add to the literature of Connected and Automated Vehicle (CAV) security by incorporating the security vulnerabilities and countermeasures of the Vehicular Ad hoc Networks (VANETs) and their access technologies. Providing knowledge to the vehicles or other entities involved with modern transportation in such networks, is the key to proper traffic management, efficient use of resources, and to improve the transportation safety, which is referred to as Intelligent Transportation System (ITS). As a critical component of the ITS, soon to be deployed VANETs, are not immune to the cyber/physical adversarial risks. Compounding VANETs and modern vehicles will allow adversaries to gain access to the in-vehicle networks and take control of vehicles remotely to use them as a target or a foothold. Extensive attention has been given to the security breaches in VANETs and in-vehicle networks in literature but there is a gap in literature to assess the security vulnerabilities associated with VANETs' access technologies. That is, in this paper we contribute to the CAV security literature in threefold. First, we synthesize the current literature in order to investigate security attacks and countermeasures on VANETs as an ad hoc network. Second, we survey security challenges that emerge from application of different VANETs' access technologies. To augment this discussion, we investigate security solutions to thwart adversaries to compromise the access technologies. Third, we provide a detailed comparison of different access technologies performance, security challenges and propound heterogeneous technologies to achieve the highest security and best performance in VANETs. These access technologies extend from DSRC, Satellite Radio, and Bluetooth to VLC and 5G. The outcome of this study is of critical importance, because of two main reasons: (1) independent studies on security of VANETs on different strata need to come together and to be covered from a whole end-to-end system perspective; (2) adversaries taking control of the VANETs' entities will compromise the safety, privacy, and security of the road users and will be followed by legal exposures, as well as data, time and monetary losses. Outcome of this study communicates VANET and its access technologies security risks to the policy makers and manufacturers who are concerned with the potential vulnerabilities of the critical transportation infrastructures.

*Index Terms–VANETs, VLC, 5G, DSRC, WAVE.*

## I. Introduction

It was only two years after the first wired car hacking that Miller and Valasak demonstrated wireless hacking on a Jeep Cherokee and decided to demonstrate part of their work to public in order to stimulate the automotive industry as well as the policy makers to take action [1]. To this point, researchers/hackers have successfully hacked a handful of vehicles and divulged some of the vehicular architecture security vulnerabilities. Elevated demand for cutting-edge technologies, fast implementation of connected/automated vehicles and affordability, are just a few reasons why rigorous security measures and policies have been overlooked. Today's vehicles are computers on wheels equipped with Wi-Fi routers, Bluetooth modules and dozens of Electrical Control Units (ECU). Even further, some foresee cars as a platform to offer on-demand technologies and services to the customers. On-demand applications could be accessible by third-party developers to the road users, in a similar fashion as the iPhone App store [2]. Such advancements could transform vehicles into personal computers but with multitudinous security vulnerabilities [3].

Starting in the 1980s, researchers started to develop the concept of vehicular wireless communication [4]. Since then, to overcome issues with road safety, road users' comfort, transportation efficiency and environmental impacts, inter-vehicular communication (IVC) development significantly increased [5]. VANET is a subclass of Mobile Ad hoc Network (MANET) that employs Dedicated Short Range Communication (DSRC) technology tuned to the 5.9 GHz frequency. Visible Light Communication (VLC) is a common alternative technology proposed to be used in VANETs which brings in certain pros and cons once applied for vehicular communications [6]. VLC is a strong candidate for this application because the bandwidth associated with it expands from 400 THz up to 790 THz which lies far away from radio frequency (RF) and contributes to solving the RF congestion problem. Other than VLC and DSRC, the potential access technologies for VANETs include Bluetooth, 5G and Satellite Radio. VANET involves vehicle-to-vehicle (V2V) –also known as inter vehicle communication (IVC)– and vehicle-to-infrastructure (V2I) communication, relying on a wireless environment [5]. Vehicular communication deals with several security challenges that need to be resolved prior to any real-

K̲B Kelarestaghi (*corresponding author*) is with the Department of Civil & Environmental Engineering, Virginia Tech, 900 N. Glebe Rd., Arlington, VA, 22203. (e-mail: kavehbk@vt.edu; phone: 202-780-8588).

M. Foruhandeh and R. Gerdes are with The Department of Electrical Engineering, Virginia Tech, 900 N Glebe Rd, Arlington, VA, 22203. (e-mail:{mfhd, rgerdes}@vt.edu).

K. Heaslip, is with the Department of Civil & Environmental Engineering, Virginia Tech, 750 Drillfield Drive, 301 D2 Patton Hall, Blacksburg, VA 24061. (e-mail: kheaslip@vt.edu).

world implementation. These challenges are due to e.g. high node mobility, irregular connectivity, large-scale environment, availability and dilemma between privacy and authentication [5, 7].

For VANET to be successfully implemented, different players would need to trust the system. They will fail to trust VANET if security vulnerabilities exist. Although errors are inevitable, VANET environment has low tolerance for error. The success of the VANET depends on individuals to trust information that is broadcasted to them. It is critical that such information cannot be eavesdropped and /or tampered by an adversary. To stop adversaries from causing any harm to the ITS critical infrastructure, the system should provide prevention strategies in addition to detection and recovery strategies. Such prevention strategies need to be built into the design from inception. A handful of reasons to secure VANET prior to implementation are proposed in [8] as: (1) design phase is the most effective stage to prevent exposure, (2) initial security consideration will eliminate those unsecure proposals, (3) blind eye to attackers will endanger the system robustness. Additional to the aforementioned reasons, in context of VANETs, security issues are involved with passenger safety which indicates how critical VANET security issues are compared to any other ad hoc network.

In real world implementations, the vulnerabilities are not limited to VANET security issues. In fact, smart vehicles which are considered as nodes in VANETSs, own internal vulnerabilities of themselves which are referred to as in-vehicle network security issues. A malicious user can always take advantage of the VANET security breaches and find a way to access the internal bus of a vehicle, usually CAN bus and as a result take control of the vehicle braking system, engine, air conditioning system and steering wheel which in fact is remotely driving the vehicle [9]. This study attempts to survey the impacts of security vulnerabilities of ITS, by exploring possible consequences of an adversary hacking the VANET system. To this aim, we draw a bridge between VANET and on-the-market vehicle security vulnerabilities, to elucidate that it is exceedingly plausible for an adversary to breach into any of them and impose dire consequences. Cyber/physical attacks will compromise safety, privacy and security of the transportation network, followed by consequences such as data, time and monetary loss [10]–[14].

This study, first, surveys the security vulnerabilities and the proposed countermeasures for VANETs as a general ad hoc network, and next investigates the security issues and challenges that vary based upon the access technology being used by VANETs. This is the first attempt to tie these two concepts together and assess the security breaches and countermeasures in VANETs as a whole end-to-end system. More importantly, this is study is the first attempt to scrutinize the pros and cons of using each technology and compare them in terms of performance and security. As an example of access technologies, the literature on 5G are massive, however most of the contributions consider performance as the main metric and overlook the critical issues of security. We are attempting to bridge the gap between security and performance as two critical evaluation metrics for deployment of VANETs. This will be followed by a brief discussion on the ramifications of a malicious behavior on road users and operators and to suggest countermeasures to mitigate such a risk. Our key contributions are summarized as:

- Surveying VANET attacks and countermeasures.
- Surveying the attacks and countermeasures that come with application of VANETs' access technologies.
- Comparing the VANETs' access technologies from the security point of view.
- Comparing the VANETs' access technologies from the performance and feasibility point of view
- Discussing the possible impacts of an adversarial activity on road users and operators.

The remainder of this paper is organized as follows: Section 2 presents basics of the VANET, Section 3 discusses VANET vulnerabilities and defense mechanisms that have been suggested to secure VANET from a general ad-hoc network perspective summarizing the interrelation between VANET vulnerabilities and current cyber/physical attacks in the vehicle arena. In addition, in Section 4 we investigate the access technologies and the pros and cons that come along each of them as well as the specific security issues and vulnerabilities that they bring along. Finally, we conclude the paper with Sections 6.

## II. VANET BASICS

### A. Communication

In 1999, United State Federal Communication Commission (US-FCC) assigned 5.850-5.925 GHz band for vehicular communication [15]. This 75 MHz of spectrum is referred to as DSRC. DSRC is IEEE 802.11-2012 standard, proposed to amend IEEE 802.11p-2011 overhead operation [16]. Later in 2006, IEEE 1690 specified standards for Wireless Access in Vehicular Environment (WAVE) application layer and message formats [16, 17] for operation in DSRC communication. DSRC has a large bandwidth and is able to use up to seven channels [17]. The platform supports communication ranges up to 1000 meters, with 100 and 300 meters for highway and urban applications, respectively [18]. Other than DSRC communication which is the most common type of access technology being discussed today, there are several strong candidates for VANETs which outperform DSRC in terms of data rate, cost, bandwidth, or coverage depending on the situation. This includes: VLC which relies on transmitting information via visible light which is ubiquitous in the ambient; Satellite Radio which is primarily intended for occupants of motor vehicles and is already implemented and brings in global coverage; Bluetooth that is all available in the modern vehicles and already used for communicating with the users or in the infotainment systems; and last, 5G which is a promising generation of wireless communication in which different technologies get to coexist in a hybrid communication network and create a highly efficient and fast system.

*B. Characteristics vs Drawbacks*

VANET is a wireless network where fixed and mobile nodes communicate with each other. Such system has characteristics such as frequent disconnections, dynamic topology and high node mobility. Each of these characteristics may result in a security vulnerability that threatens the system. Compared to the MANET, VANET has relatively higher mobility. In VANET, vehicles move randomly within the network and their movement is constrained to the network topology. Due to the high node mobility, VANET topology is dynamic and unpredictable; due to this and other factors such as adverse weather condition, VANET may suffer from various disconnections throughout the system. Other characteristics of the VANET are indicated in Table 1.

**Table 1** VANET's characteristics [19].

| Characteristic | Description |
| --- | --- |
| High node mobility | VANET has higher mobility compare to MANET, which reduces the mesh in the network (possible routes for communication) [20], [21]. |
| Dynamic topology | Topology of the network alter fast, the durations of the communications are short and density of nodes vary extensively based on the location and time of the day which can facilitate cyberattacks [22]. |
| Frequent disconnections | Adverse weather conditions in addition to the network mobility, high speed of nodes, and topology may cause frequent disconnection [22]. |
| Transmission medium | Air is a transmission medium which is accessible to adversaries [22]. |
| Anonymity of the support | The anonymity structure might allow adversaries to act maliciously within similar frequency band [8]. |
| Energy storage and computing | No energy constraints. However, the real time environment is a challenge [22]. |
| Predictability | The dissemination of nodes is limited by the city roadmaps which makes the routing of nodes more predictable [19] |

*C. Model*

VANET model is structured into infrastructure and Ad-hoc environment [23]. In the infrastructure environment, different entities can be interconnected. These entities include the following: manufacturer, Trusted Third Parties (TTPs), legal authority and service provider. Manufacturer may be responsible for vehicles' unique identification, while legal authorities are responsible for vehicle registration and offence reporting. TTPs and service providers both are offering different services in VANET environment such as credential management and location-based services (LBS). On the other hand, the ad-hoc environment is where V2I and V2V communications are taking place. This communication occurs through OBU, sensors and Trusted Platform Module (TPM)– TPMs are installed for security purposes [23]. Depending on the access technology being used, the connection between the infrastructure and the ad-hoc environment varies. For example, this connection happens via RSU if the access technology is DSRC, or via the 5G BTS of the access technology is 5G. Section 4 includes multiple figures that illustrate the VANET model and how it changes based on the access technology.

Figure 1 illustrates a simple structure for the VANET model.

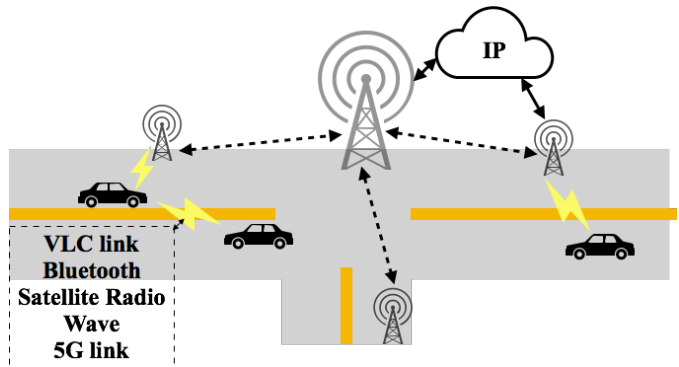

**Figure 1** Architecture of VANETs.

*D. Application*

VANET's application setting can be classified into four main sections [23]; V2V warning propagation, V2V group communication, V2V beaconing, and infrastructure-to-vehicle/vehicle-to-infrastructure (I2V/V2I) warning. Applications that gain the most attention can fit into safety-related and comfort-related applications [24]–[27]. On the latest fact sheet provided by US Department of Transportation [28], twenty eight safety applications on V2I and V2V environment have been introduced; *pedestrian in signalized crosswalk warning and pre-crash actions* are among those applications. On the other hand, non-safety applications aim to provide comfort for road users in addition to improve traffic efficiency (e.g. eco-traveler information, eco-lanes management). The complete list of applications can be found in [29].

### III. VANET SECURITY

*A. Adversarial categories*

Acquiring a solid understanding of the nature of a potential adversary and its capabilities is essential in order to provide a secure communication network of any type. Based on membership, motivation, method and scope, an attacker can lie into four categories [24, 25] in the context of VANETs: (1) Insider (legitimate member of a system) and outsider (intruder with more limitation than insider). (2) Malicious attacker with no personal profit (e.g. pranksters) and rational attacker who pursues reward but, may be more predictable than malicious attacker (e.g. greedy driver). (3) Active attacker to generate packets and passive attacker with aim of eavesdropping (e.g. snoops). Depending on how sophisticated an adversary is, his/her attack may result in dire consequences. In spite of other type of attackers, insiders (or industrial insider) may impose highest risk to the VANET, yet the likelihood is very low and can be prevented. (4) Local attacker has limited control over entities, while an extended attacker is able to control various entities throughout the VANET in a larger extent. Depending on how sophisticated an adversary is, their attack may result in dire consequences. Unlike other type of attackers, insiders (or industrial insider) may impose highest risk to the VANET, yet the likelihood is very low and can be prevented.

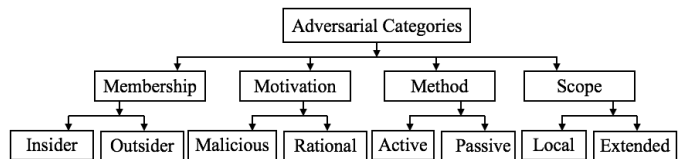

**Figure 2** Adversarial categories for VANETs.

## B. Challenges

Technical and socio-economic challenges are the key challenges of VANETs; the first one deals with system functionality as well as challenges to balance security and privacy needs, while the later deals with market introduction and penetration [5]. Although socio-economic challenges may impact different parties (e.g. road operators), technical issues are considered a fundamental constraint to a successful deployment [26, 27]. This means that even at the presence of socio-economic assurance, solutions for technical challenges should be available [31]. To maintain networks of different types in general, a protocol should address the following technical requirements: (1) authentication, (2) confidentiality, (3) availability, (4) integrity, (5) non-repudiation, and (6) privacy. Deployment of VANETs however, not only mandates the aforementioned requirements on security perspective to be satisfied, it imposes strict throughput requirements as a highly dynamic real-time system, and faces VANET specific challenges like key distributions at a short period of interaction between the vehicles due to mobility of system. Last not least, bringing in a new technology is a smooth transition, proving the benefits of VANETs in a smaller scale is a non-trivial task for the manufacturers due to limited number of users. Negligible as it might seem, the latter highly effects the legislators on one hand and decision makers at technical level on the other hand. The following explains how the general network security requirements match to VANETs.

- *Authentication:* all nodes within the VANET have to be authenticated to be secure from an attacker that is trying to compromise the system using a falsified identity. Failure to maintain a proper authentication system can result in drastic degradation in a communication network [34, 35].
- *Confidentiality:* which is another requirement that assures only a legitimate user can access the data [21]. Failing to address confidentiality puts the security of the exchanged data at stake and affects privacy of the users [23].
- *Integrity:* the guarantee that the message content remains secure. Attacks on integrity and authentication will enable adversaries to inject false messages [36]. Preserving data integrity is one of the most important VANET security goals. The message content should be secure while communicating V2V and V2I [37], and adversaries should not be able to alter or fabricate the messages.
- *Availability:* it is an important requirement of the VANET that guarantees system functionality at all time. Although attacks on availability have been selected as the most critical attacks [37], in [38] authors believe that data integrity and confidentiality have higher priority than availability. This is due to the risks that are associated with sensitive information being disclosed by adversaries. Since safety messages will not include any sensitive information, authentication will be sufficient to provide security for safety messages (no need for encryption) [31].
- Non-repudiation security goal should be addressed [17, 31]. Satisfaction of the non-repudiation requirements assures that an entity cannot deny having sent or received a message [40].
- *Privacy:* another challenge of VANET is that the privacy of users should be maintained. A successful protocol should be able to preserve drivers' privacy by preventing adversaries to track their vehicles and to link vehicle identity with the driver's identity [23]. However, this security goal should not prevent trusted authorities to identify drivers or track vehicles. To solve this issue, Lin et al. [41] introduced secure and conditional privacy preservation protocol (GSIS). This protocol found to preserve the privacy of the entity while providing trusted authorities' ability to trace a vehicle in any traffic event dispute.

## C. Attacks & Countermeasures

Reliance of ITS on IVC with ordinary nodes for a decentralized operation (unlike the conventional counterparts which is controlled by a central location such as the railroad dispatch system) makes it more vulnerable to cyberattacks and may result in an intelligent collision [8]. To thwart attacks such as Denial of Service (DoS), Blum and Eskandarian proposed secure communication architecture called SecCar [8]. This virtual infrastructure will assure IVC availability and confidentiality, and will identify the malicious users' identities. Also, with the non-repudiation feature, entities cannot deny participating in an attack. Parno and Perrig [30] proposed authenticated localization of message origin to address secure relative localization in VANET which prevents attacks such as wormhole attacks which is a challenging attack to defend against because in this type of attack all of the nodes in the network can provide authenticity and confidentiality. In brief, the attacker records packets in one location of the network and tunnels them to another location in the network [42]. In their proposed scheme, anonymization service, provides a solution for conflict between authentication and privacy. Drivers' short-term identity simultaneously, creates secure authentication and prevents adversaries from spoofing identities. In addition, use of secure aggregation techniques for applications and primitives for message authentication and key establishment are suggested.

Raya et al. [7] proposed security architecture that comprises of different components: (1) Event Data Recorder (EDR), (2) Tamper-Proof Device (TPD), (3) Vehicular Public Key Infrastructure (VPKI) based on (4) Elliptic Curve Cryptography (ECC). EDR acts like an airplane black box. EDR records vehicle's critical information such as location and speed, while TPD is responsible for cryptographic operation and storing cryptographic information, particularly signing and validating safety communications. Certificate Authorities (CAs) will issue VPKI for vehicles. CAs shall be cross-certificate so vehicles of different regions can authenticate each other. To secure authenticity, vehicles need to sign messages with their keys and attach the corresponding certificate. The security architecture proposed the use of ECC to manage security overhead. To address privacy issues, anonymous keys that changes frequently by speed may be used (every couple of minutes); the keys will be preloaded in TPD for a certain period of time (1 year). In spite of addressing basic requirements of a secure VANET, there are certain shortcomings in [7] against jamming GPS signals for instance that leads to DoS and severe consequences. The signals are weak and vulnerable to jamming, can be spoofed easy and simple approach of frequency hopping is no more able to solve this problem.

The following section provides, survey of threats to VANET's availability, authentication, identification, confidentiality, integrity, data trust, privacy and non-repudiation, along with defense mechanisms to prevent them.

*D. Availability*

Availability assures that VANET is functional. Threats to availability can be directed towards both communication network and participating nodes [23]. There are different types of attacks that can be launched to bring down the availability of VANETs. Previous studies have proposed various security mechanisms to thwart threats on the availability.

Among attacks that threaten the availability of the VANET, DoS is the most famous one that can affect both RSU and OBU to jam and compromise the network. In DoS attack, adversaries aim to either overload the communication channel or to create interferences to hinder the use of the communication channel. Signature based authentication scheme, proposed by He and Zhu [43], based on the pre-authentication process, reduces the impact of DoS attacks. This scheme prevents attackers from forcing the receiving entities to perform redundant signature verifications.

Distributed DoS (DDOS) also can threaten VANET availability by sending frequent transmissions where more than one entity is sending out malicious messages [44]. Biswas et al. [44] illustrated that VANET with IEEE 802.11p mechanism is vulnerable to the synchronization-based DDoS attack mainly caused by small contention window sizes and periodicity of service beacons, and it is possible that neither sender or receiver notice the attack, due to lack of acknowledgement mechanism. Authors put forth a solution to address this problem which require modifications to MAC layer mechanisms. Increasing the contention window size, and randomization of inter arrival time of service beacons are the basis of their solution to prevent DDoS attacks. In the case of timing attack, adversary is trying to delay the messages that need to be broadcasted to an adjacent vehicle. To mitigate this attack, Mejri et al. [22] proposed a cryptographic solution by providing the timestamping mechanism to certify the time of an event.

Group-signature based schemes are also beneficial in the case of message tampering or alteration. In this scheme, 1) each member of a group can anonymously sign messages, and 2) a group manager can trace a signer [45]. To preserve the privacy of the group members, unlinkable schemes were proposed (e.g. [46]), in which a user's various interactions in the VANET cannot be linked [47]; a verifier cannot determine whether the messages are being transmitted from the same vehicle or different vehicles. This makes unlinkable group signature based schemes vulnerable to the Sybil attack [45]. To solve this issue, Domingo-Ferrer and Wu [45] proposed message-linkable group signatures (MLGS) scheme. With MLGS it is easy to identify whether two signatures on one message are generated by one entity or two entities. In general, Sybil attacks are on top of the security threats to VANETs.

Wormhole is another DoS attack that can compromise the VANET routing protocol. This attack is executed by creating a tunnel between two or more malicious entities to transmit data packets [31]. This type of attack can be launched without prior knowledge of the network. Those malicious entities can execute DoS attack if gaining unauthorized access [48]. Hu et al. [49] proposed Packet Leashes mechanism to limit the maximum distance that a packet can be transmitted, hence prevent the tunneling operation.

Another attack that can threaten the availability of VANET is the black hole attack. In this attack, a node either declines participation in a routing operation or drops out data packets. Dropping out can cause a message propagation failure [21]. Cherkaoui et al. [50] proposed a cluster-based algorithm that provides security scheme to thwart the black hole attack. The algorithm identifies the attack and isolates the malicious entity. They used the Simulation of Urban Mobility (SUMO) traffic simulation to simulate a real-world mobility model and to execute the black hole attack. In this study Cluster-Head oversees V2V communications within the cluster and is responsible for detecting black hole attacks and to broadcast warning messages to other members of the cluster. Gray hole attack also is similar to the black hole, where an adversary focuses on applications that are sensitive to the packets' loss [22]; In black hole attack a malicious node drops all data packets while in the gray hole attack the node always forwards its own packets [51]. In this attack, the behavior of the attacker is random, since the node can act maliciously for specific periods of time while other times it can be as normal as other nodes [52]. Survivable Ad hoc and Mesh Network Architecture (SAMNAR), proposed by Nogueira et al. [51], provides reactive and tolerant security mechanism to thwart DoS attacks such as gray hole attacks. The result of simulation indicates that SAMNAR can significantly improve the security to minimize the performance loss.

Greedy behavior is another type of attack where a greedy node does not follow the waiting time regulation for accessing the channel based on CSMA/CA protocol. Upper bound enforcement on the access time of nodes after a training process is the solution proposed at [53].

Several other attacks that threaten VANET availability are Malware (i.e. viruses and spam) [48, 54], spamming [21, 54], and jamming [55]. Table 2 indicates a list of attacks and security mechanisms to prevent those attacks.

*E. Authentication, Identification and Privacy*

In VANET each entity should have its own identity that is called entity identification [23]. The requirement to verify the rightfulness of this identification is entity authentication [23]. For VANET, authentication includes user authentication and message integrity [39]; entities should only reply to messages that are coming from an authorized source [48]. Adversaries can threaten the authenticity of the system by impersonating vehicles or RSUs. A malicious entity can transmit false messages to make-believe that a certain road is congested, to dissuade others from using that road [21].

To preserve authentication, a unique identifier should be assigned to each entity. This identification does not need to be the actual identity for each of the participating entities [23]. Similar to the concept of license plate, Hubaux et al. [56] proposed Electronic License Plate (ELP) that allows a vehicle to be both identified and authenticated. ELP certifies vehicle identity and can be used for dynamic pricing and to identify vehicles that escape a scene of a crash [56]. However, since ELP is linked with entity real identity, public key certificates are proposed as a substitute [23]. VPKI is designed to assure that public key belongs to an authenticated node. In this scheme, ELP is linked with a pseudonym that is being managed by the VPKI, and a legal authority can link the pseudonym with the owner's real identity [23].

To secure authentication and identification, Gollan and Meinel [57] proposed a combination of digital signature and Global Positioning System (GPS) to substitute traditional solutions (e.g. license plate) that are easily forgeable. GPS signals however are weak and vulnerable to spoofing. There are different countermeasures for encountering GPS spoofing at [58]. Zeadally et al. [21] identified digital signatures as a suitable mechanism to thwart attacks on authentication. Raya and Hubaux [31] used digital signature as the most efficient approach. With this method, a sender digitally signs the message to address the authentication requirement.

**Table 2** Availability - Attacks and Security Mechanisms.

| Security Goal | Cyber Attack | Security Mechanism |
|---|---|---|
| Availability [31], [59], [23], [21], [60] | DOS [20], [21], [61] <br> DDOS [44] <br> Jamming [62] <br> Greedy behavior [63] <br> Broadcast tampering [21] <br> Malware [20], [48] Spamming [20], [21] <br> Message Fabrication/alteration/tampering/suppression [22] <br> Black hole [20], [21], [48] <br> Gray hole [22], [51], [52] <br> Sinkhole [22] <br> Wormhole [64], [65] <br> Tunneling [66] <br> Timing attack [48], [67] <br> Sybil attack [66] | Signature based authentication mechanisms [43], [31], [45] <br> Secure Communication Architecture (SecCar) [68] <br> MAC layer modifications on contention window size, and randomization of inter arrival time of service beacons [44] <br> Switch between channels or even communication technologies. In worst case scenario VANET should turned off [31]. Frequency hopping technique FHSS consist of cryptographic algorithm [22] <br> CSMA/CA modifications [63] <br> Timestamping mechanism [21] <br> Message-linkable group signatures (MLGS) scheme [45] <br> Cluster-based algorithms[51] <br> SAMNAR [51] <br> SAMNAR [51] <br> Packet Leashes mechanism [49] <br> Packet Leashes mechanism [49] <br> Cryptographic solution using timestamps [21] <br> Message-linkable group signatures (MLGS) scheme [45] |

A well-known threat to authentication is Man-in-The-Middle (MITM) attack, wherein malicious node eavesdrop the communications and inject false data [48]. To secure the system from MITM attacks, authentication can be provided with VPKI [8]. Hesham et al. proposed using both ELP and Chassis number to create Vehicle Authentication Code (VAC) to support PKI-based VANET [69]. Distance bounding protocols are also proposed to secure the authentication especially against MITM and wormhole attacks [70].

Hubaux et al. [56] proposed distance-bounding protocol as a solution for authenticating vehicle location; their protocol can also detect jamming attacks. Distance-bounding protocol aims to provide upper bound for a distance between authorized nodes, by timing the delay between transmission and response [71]. Illusion attack is another attack that threatens the authentication of the system. In the case of an Illusion attack, an adversary transmits fake traffic warning messages to an adjacent vehicles and can cause illusion situation (e.g. traffic congestion) [40]. While traditional message authentication and integrity check are abortive to thwart the illusion attack [40], Plausibility Validation Network (PVN) stated to be a successful approach [72]. PVN conducts plausibility check on the collected data received by the vehicle, and examines the validation of the messages [72].

Securing authenticity and privacy of the system concurrently, causes conflict [5]. For instance, the conflict can transpire when an unauthorized user takes advantage of VANET protocol and gains access to the location of a vehicle [73]. While it is critical to secure the system, privacy of the users have to be maintained [31]. Several methods have been proposed to overcome this problem. Application of a random/temporary identifier is proposed as one of the main strategies to overcome this issue [74]. Sampigethaya et al. [73] investigated an anonymity-based mechanism (AMOEBA scheme) to protect the system against location tracking. AMOEBA applies the concept of group navigation of vehicles (cluster neighboring vehicles in groups) to attain location privacy.

In tracking attack, the location of a vehicle in a certain time can be captured by an adversary [75]. In [7] and [31] anonymity techniques are proposed to secure the privacy of the vehicle. Both studies use anonymous key pairs (public/private key pair) to protect privacy. This key is authenticated by the certification authorities (CA) and thwarts tracking attack. To protect driver's identity and to minimize the storage cost, the key changes according to the vehicle speed to ban observer from tracking the key owner.

Group Signature and Identity based Signature (GSIS) is a cryptographic approach presented by Lin et al. [41], aimed to tackle the tradeoff between security and privacy preservation. GSIS benefits from both group signature and identity-based signature, to provide security and privacy for V2V and V2I communications. In this scheme group signature assures the anonymity of the sender, and the identity-based signature assures the authenticity of a message. This method is however time consuming and might not be the best choice for VANETs [76]. Freudiger et al. [77], proposed a creation of mix-zones (anonymizing regions), by use of cryptography, to address drivers' location privacy. This technique provides security by preventing adversaries from gaining access to the messages and preserving the privacy of drivers by changing pseudonyms (to ensure unlinkability).

In addition to the listed attacks that threaten the authentication and identification of a system, there are several other approaches such as a naive Brute force attack. This type of attack can be easily avoided by using large enough keys and identification numbers that require huge amount of time to break [78]. Table 3 provides a survey of cyber threats and several defense mechanisms to secure VANET authentication, Identification and privacy.

### F. Confidentiality, Integrity and Data Trust

Confidentiality in the VANET environments assures that messages are only being read by legitimate node, while Integrity and data trust aims to protect messages from manipulation and alteration [23]. Insider and/or outsider adversaries can threaten the confidentiality of the exchanged messages by eavesdropping and information gathering attacks (passive attack) [21]. Data trust can be compromised in a case when an adversary manipulates a message to for instance sends a false warning or inaccurate information to affect VANET reliability. To safeguard communications confidentiality, and integrity and data trust, security mechanisms should be built-in-design.

**Table 3** Authentication, Identification and Privacy - Attacks and Security Mechanisms.

| Security Goals | Cyber Attack | Security Mechanism |
|---|---|---|
| Authenticity and Identification [15, 25], [39] | Eavesdropping [66] <br> MITM [48], [79] <br> Brute force [22] <br> Illusion attack [40] <br> DOS [80] <br> Sybil attack [80] <br> Replay [30] <br> GPS spoofing/position faking [14, 35] <br> Message Fabrication/alteration/tampering/suppression/deleting [48] <br> Masquerading [48] <br> Tunneling [15, 66] <br> Key and/or certificate replication [66] <br> Node impersonation [48] <br> Black hole [66] <br> Grey hole [66] <br> Wormhole [48] <br> Message saturation [66] | Vehicle Authentication Codes (VAC) [64] <br> Augmented digital signatures and GPS data [53] <br> Authorized access on hardware and software [22] <br> Access to sensors should be authenticated and verified [22] <br> Deployment of trusted hardware [82] <br> Plausibility Validation Network [72] <br> Secure Location verification [6, 68] <br> Timestamping mechanism [21] <br> Digital Signatures [43], [56], [84], [85], [86], [87], [88], [8], [31], [45], [89], [7], [23] <br> Secure Communication Architecture (SecCar) [8] <br> Validation check with CRL (Certificate Revocation List) [31] <br> Variables MAC and IP addresses [31], [7] <br> Distance Bounding protocols [56], [71], [82] <br> Electronic License Plate (ELP) [56] <br> Timed Efficient Stream Loss-tolerant Authentication (TESLA) [90] |
| Privacy [31] | Eavesdropping[66] <br> Tracking [66] <br> Social engineering [79] <br> Tampering hardware [66] <br> GPS spoofing/position faking [66], [91] | Digital Signatures [21], [31], [40], [92] <br> Augmented digital signatures and GPS data [69] <br> Variables MAC and IP addresses [31] <br> Anonymity techniques [38], such as key changing algorithm [31] <br> Use of anonymous, and preloaded keys that change according to the vehicle speed [7] <br> Use of random identifiers that change over time [74] <br> Location Cloaking techniques [93] <br> Data Aggregation Mechanism [94] <br> AMOEBA [73] <br> Group signature based scheme: GSIS [41] <br> Silent Period [95] <br> Mix-zones [77] <br> Controlling manufacturer employees [66] <br> Timing based sanity checks for GPS protocol [58] |

Besides authentication and identification, the Brute force attack is also against the confidentiality of the VANET. In this attack an adversary aims to find the cryptographic key or the vehicle network ID [22, 96]; since the duration of communications are typically short, executing such attack can be challenging. Langley et al. [97] proposed large random value as a vehicle's VIN to thwart Brute force attack. For an attacker to assure guessing a correct VIN combination he needs to determine this random value as well. This security posture includes additional security measure, to decreases the likelihood that an attacker can compromise the vehicle identity.

Other mechanisms to secure VANET confidentiality are Secure Group Communication (SeGCom) [98] and Situation-Aware Trust (SAT) [99]. In SeGCom scheme, proposed by Verma and Huang [98], RSUs are responsible to establish a symmetric key (for a period of time) with vehicles that are entering the region. To increase the accuracy of group formation, the region separated into *splits* with split key, to be used for group communication. Hong et al. [99] proposed a descriptive attribute-based encryption to improve entity trust among vehicles. Attributes can be used to identify the common properties for a group of vehicles and enforce data access policies accordingly. In this approach, only group members are able to decrypt the transmitted messages. SAT model is capable of thwarting attacks such as eavesdropping, tracking, forging and jamming [100].

MITM and Illusion attacks are among the attacks that are threatening the integrity and data trust of the VANET [40, 48,

79]. In the MITM attack an attacker actively eavesdrop the communication between vehicles and eventually attempts to inject fabricated messages between those vehicles [48, 79]. Attacker intention is to control the communication while the vehicles presume otherwise. In the illusion attack adversaries use sensors to inject false data, such tampering would have an impact on drivers' behavior [40]. An adversary can sway the traffic to avoid a certain road by falsifying that there is a heavy congestion along that road. Authentication protocols may have minimal capability to thwart illusion attacks since an adversary connection is genuine [22]. In contrary, the replay attack can formed by an unauthorized node [22] and it can be prevented by applying timestamp mechanism [21]. In the replay attack an adversary can repeat or delay the transmission of messages in order to manipulate the data, hence, violate the integrity of the data.

Masquerading attack is another threat against VANET integrity. In this attack, an attacker appeared as an authenticated node and is able to inject false messages. Raya and Hubaux [31] stated that digital certificates are effective to overcome this attack. In this approach, all nodes need to sign their messages with their public keys, and such key is unique and verifiable. To thwart the message fabrication attack, VPKI and zero-knowledge –an entity proves its authenticity without revealing its identity– techniques can assure the authenticity of a node and validity of the message [22, 74]. Similar approach can also prevent MITM attacks, by providing robust authentication [22]. Park and Zou [101] proposed two-directional verification technique to secure data integrity in VANET. With this approach recipient entity receives a message twice and if one of them is false the entity would not assent either of them. And if only one message being received by the entity, the message can be approved with low confidence.

Brute force attacks, cryptographic information replications and tampering with hardware are other possible attacks to be named here, threatening the confidentiality and integrity in VANETs. Table 4 indicates a list of threats and security mechanism in terms of confidentiality and data integrity of the VANET.

**Table 4** Confidentiality, Integrity and Data Trust - Attacks and Security Mechanisms.

| Security Goals | Cyber Attack | Security Mechanism |
|---|---|---|
| Confidentiality [59], [20], [21] | MITM [48], [79]<br>Brute force [22][96][78]<br>Eavesdropping [20][22]<br>Traffic analysis [78]<br>Tampering hardware [66]<br>Key and/or certificate replication [66] | Encrypt data with great importance, Application of VIPER (Vehicle-to-Infrastructure communication Privacy Enforcement Protocol) algorithm [102]<br>SAT (Situation-Aware Trust) [99]<br>Group key management protocol (GKMP) involve with geographically defined group [103], [104]<br>SeGCom (Secure Group Communication) [98]<br>Controlling manufacturer employees to prevent tampering with hardware [66] |
| Integrity and Data Trust [59], [105], [106], [23] | MITM [48], [79]<br>Illusion attack [40]<br>Message suppression/fabrication/ alteration/tampering [81], [30]<br>Masquerading [22]<br>Replay [30]<br>Black hole [21], [48], [54]<br>Gray hole [51], [52], [107]<br>Worm hole [51], [48], [64], [65]<br>Message saturation [66]<br>Broadcast tampering [66]<br>Node impersonation [66] | Digital certificates [7], [23]<br>Authorized access on hardware and software [22]<br>Access to sensors should be authorized and verified [22]<br>Deployment of trusted hardware [82]<br>VPKI [8], [31], [45]<br>Group communication in which group key management system (GKM) is responsible for keys [45], [75]<br>Secure Communication Architecture (SecCar) [8]<br>Timestamping mechanism [21]<br>Plausibility Check [72]<br>Two directions Reporting [101]<br>Threshold-based trust [47]<br>MLGS (Message Linkable Group Signature) [45] |

## G. Non-repudiation

VANET Non-repudiation goal is to assure that different entities are not able to deny their actions. In repudiation-type attacks several entities use similar identifications to repudiate their actions [23]. For instance, a vehicle in the VANET can deny sending a message (repudiation of origin) or either receiving one (repudiation of receipt). Non-repudiation of origin (NRO) and receipts (NRR) assure that an entity is accountable for the messages that it transmitted and received.

Elliptic Curves Cryptography (ECC) proposed among the best mechanisms to ensure NRO [23]. Basically, IEEE 1609.2 standard [108] supports Elliptic Curve Digital Signature Algorithm (ECDSA) because, compared to other algorithms ECDSA provides smallest keys and signatures. As a subtype of digital signature application of the group signature mechanisms ensures the NRO [109, 110]. Although group signature based mechanisms are apt to ensure NRO, there are some glitches when it comes to ensuring the NRR [23]; for instance, having a group leader in between of the RSU and the group members can threaten the availability of the group leader.

**Table 5** Non-Repudiation – Attack and Security Mechanisms.

| Security Goals | Cyber attack | Security Mechanism |
|---|---|---|
| Non-Repudiation /Accountability [31][39] | Masquerading [31]<br>Loss of event traceability [22]<br>MITM [66]<br>Message suppression/fabrication/ alteration/tampering [66]<br>Node impersonation [66]<br>Replay [66] | Electronic License Plate (ELP) [56]<br>Secure Communication Architecture (SecCar) [8]<br>Authorized access on hardware and software [22]<br>Access to sensors should be authorized and verified [22]<br>Deployment of trusted hardware [82]<br>Elliptic Curves Cryptography (ECC) [23]<br>Signature based schemes [109], [110], [8], [31], [45] |

SecCar scheme, proposed by Blum and Eskandarian [8], also ensures the non-repudiation of the VANET. With distributed intrusion detection system, entities can reject forged messages and enforce non-repudiation. To thwart loss of traceability (repudiation), Mejri et al. [22] suggested only an authorized access to the hardware and software in the VANET. Singelee and Preneel [82] suggested the deployment of a trusted hardware in which an adversary cannot alter the existing protocols and values. Summary of threats and security mechanisms in terms of non-repudiation are listed in Table 5.

Figure 3 summarizes the threats and security measures in terms of VANET six different types of general security goals. The connector links are color coded according to each of the security goals. For instance, the availability goal is color coded as red. In addition, the links are placed in order, initiated from the availability goal (color coded in red) and ended with privacy goal (color coded in black).

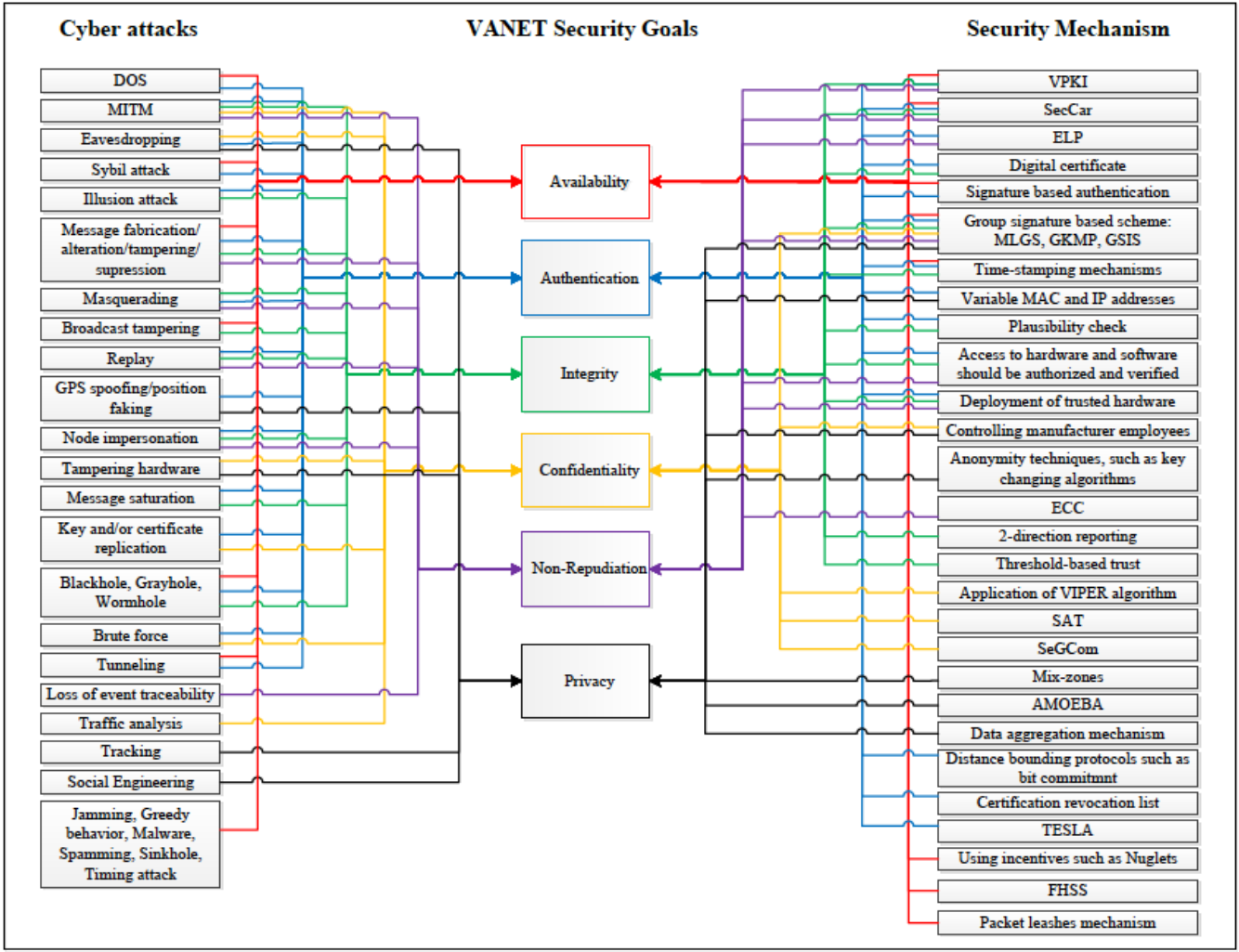

**Figure 3** Schematic view of VANET cyberattacks and security mechanisms.

## IV. SECURITY ISSUES BASED ON ACCESS TECHNOLOGY

Several access technologies have been proposed for VANETs in the state of the art. Security vulnerabilities vary based on the technology being used, hence, we overview the potential technologies for implementing VANETs as well as the security issues that come along with them.

### A. WAVE (802.11 p):

The standard is the improved version of IEEE 802.11a with improvements in MAC layer and physical layer in order to enhance the robustness in highly mobile vehicular environments. On the physical layer, OFDM modulation is applied on 7 channels of 10 MHz which are further divided into 52 subchannels. The messages are categorized into four groups based on their priority and CSMA/CA is used as the collision avoidance mechanism in a half-duplex communication which can provide up to 27 Mb/s with a coverage of up to 1000 meters. However, the NLOS communication path leads to high delays and packet collisions which reduces the reliability [6].

WAVE has three main components: Application Unit (AU), Onboard Unit (OBU) and Roadside Unit (RSU). AU can communicate with OBU through wire or wireless and can be located in the same physical unit of the OBU [12, 13]. OBU, other than providing AU a communication access, exchanges data with RSUs and other OBUs. Some of OBU's major functions are wireless radio access, network congestion control, data security and IP mobility [112]. On the other hand, RSU can extend the range of ad hoc network communication, run safety applications and also providing internet connectivity to OBUs [111]. Just like any other technology, DSRC faces certain limitations which mostly stem from the highly dynamic, fast moving, and non-predictable networks some of which are given at Table 6.

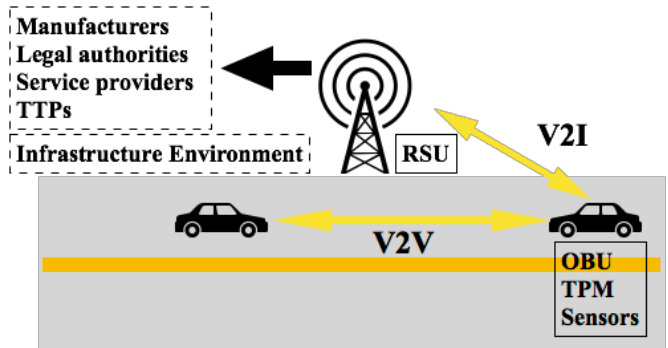

**Figure 4** DSRC as the access technology for VANETs.

**Table 6** DSRC's limitations.

| Characteristic | Description |
| --- | --- |
| Limited bandwidth | 75 MHz of DSRC band barricade its application in some countries [22] |
| Attenuations | Shortcomings of DSRC's transmission cause issues due to frequencies, fading, Doppler effect and delays [22]. |
| Limited transmission power | Maximum transmission distance in most cases is 1000 meters where *repeaters* as an intermediate device can be implemented [22]. |
| Nonreliable in highly dynamic environment | Vehicles move randomly within the network and their movement is constrained to the network topology. |
| Latency | Due to the high node mobility, VANET topology is dynamic and unpredictable; due to this and other factors such as adverse weather condition, VANET may suffer from various disconnections throughout the system. |

*B. Bluetooth IEEE 802.15.1:*

Bluetooth is a widely used standard for shortrange communications. It was first invented by Ericsson to replace RS-232 cable connections [113, p. 15]. Developed by Jim Kardach, Bluetooth works conventionally between 2402 MHz and 2480 MHz which contains 79 channels of 1 MHz bandwidth each [114]. Since 2.4 GHz ISM is already a congested band, Bluetooth uses a frequency hopping (FH) technique with rate of 1600 hops/sec in order to avoid congestion with WiFi for instance. A lower power protocol named Bluetooth Low Energy (BLE) however, uses only 40 channels of 2 MHz, among which three channels are used for broadcasting purposes and the rest are used for data transmission. This technology supports connection speed of up to 24 Mbps (Bluetooth 3) and is commonly used in intra-vehicle applications in the infotainment system, navigating via GPS and allowing phone calls via in-car interface. Easy deployment and the fact that the technology is already recognized, make Bluetooth a good candidate for VANETs. On the other hand, the proper operating range for Bluetooth is 10 meters while it extends up to 100 meters at the price of higher complexity or lower quality of service. Whether or not this operating range is suitable targeting applications in VANETs is discussed in a framework [115], where authors justify the applicability of BLE for VANETs by exploring the range, signal quality and latency of BLE signals in vehicular environment. They conduct experiments using smartphones equipped with BLE attached to cars and read the quality of channel in different environments. They compare BLE with WiFi and DSRC and conclude that BLE is a strong candidate for VANETs in dense areas. To be more specific, BLE has already been implemented (unlike DSRC) and has high power efficiency (unlike WiFi) and its limitation is mainly the range which is investigated in detail in their work. The performance of the system such as latency and throughput are important metrics however are beyond the scope of this work. We cover the security aspect of using Bluetooth for VANET communications.

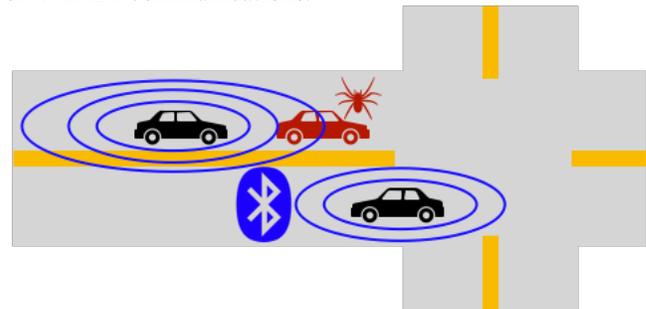

**Figure 5** Bluetooth as the access technology for VANETs.

New security challenges come along however if Bluetooth is used as the main access technology in VANETs. An open source Bluetooth packet sniffing hardware and software named, Ubertooth was designed by Michael Ossmann, which is the most affordable (130$) Bluetooth testing and monitoring platform [110, p. 7]. Considering that the technology is already on the market, this can be an indication of a well-recognized security vulnerability in Bluetooth transmission. D. Spill et al for instance demonstrate a packet sniffing by reverse engineering the hopping pattern of the FH system at a Bluetooth transceiver by extracting the MAC address (MAC address is used for generating the pseudo random hopping pattern in Bluetooth protocol) of the device in a framework that they call Bluesniff [117]. Bluesniff uses Ubertooth as the main hardware and software platform. There are several demonstrations of attacks on Bluetooth to be named here such as eavesdropping the pairing pin [118], or performing a known plaintext attack on the Bluetooth keystream generator demonstrated at [119] and extended at [120].

The current Bluetooth standard targets limited security services such as authentication (a challenge-response scheme [121]), confidentiality, authorization, integrity, pairing or bounding in a far from perfect way. Other services such as non-repudiation are missing parts of this patch that need to be implemented.

There are 4 modes in the standard that define when the security mechanism needs to be initiated, named mode 1-4 (the Bluetooth 2.0 supports only modes 1-3) summarized in Table 7. The security procedures that are initiated at service level, are basically enforced after link establishment but before logical channel establishment. In fact, the authentication and encryption mechanisms are implemented in the controller (there is a centralized security manager that maintains policies in the Bluetooth architecture). Security Mode 3 mandates authentication and encryption for all connections to and from the device before the physical link is fully established. This mode lacks authorization and can lead to security vulnerability named "authentication abuse." Authorization is basically allowing the control of resources by making sure the device is authorized to do so. In this mode, since authorization check is not performed, an authenticated remote device can access a Bluetooth service without the device owner's knowledge which is called "authentication abuse". Last, not least, mode 4 enforces service level security mechanisms. This mode has 5 levels of security which are shown in Table 8. It uses secure simple pairing (SSP) and provides authenticated link keys as well as secure connections if implemented on level 4 which is the strongest one in terms of security. Looking into Table 7 the question is "why mode 1 still exists?" Due to backward compatibility! (should be eliminated from VANETs because the risk impact is really high if devices are tricked to switch to this mode) [122].

**Table 7** Security modes in Bluetooth.

| Security Mode | When the device initiates security |
|---|---|
| 4 | Service |
| 3 | Link |
| 2 | Service (2.0 and earlier**) |
| 1 | Never (2.0 and earlier**) |

According to NIST guide to Bluetooth security [122], the number of vulnerabilities in Bluetooth implementations and specifications which provide the platform for attacks on Bluetooth can be summarized as per below. We limit the discussions to Bluetooth 4.0 and above due to lack of space as well as to maintain focus on more recent technology.

**Table 8** Mode 4 security levels, protections and specifications.

| Mode 4 levels | Security Requirements | Provides MITM protection | Encryption Required |
|---|---|---|---|
| Level 4 | Authenticated link key using Secure Connections required | Y | Y |
| Level 3 | Authenticated link key required | Y | Y |
| Level 2 | Un-Authenticated link key required | N | Y |
| Level 1 | No security required | N | Y |
| Level 0 | No security | N | N |

1. Just works association model (in this mode at least one of the devices do not have the required hardware for performing the authentication, hence there will be no authentication) opens the chances for unauthenticated link keys and is vulnerable to MITM attack.
2. SSP Elliptic Curve Diffie-Hellman (ECDH) [123] keys are weakly generated, non-unique or static.
3. Security mode 4 devices or units are allowed to fall back to all previous modes including mode 1 which facilitates even simple attacks.
4. Unlimited authentication requests can be made. Even though there is an exponential waiting time between authentication requests so as to prevent brute force attacks on the keys; such waiting times do not exist for the challenge-response requests. Hence, an attacker can brute force the challenge responses and consequently figure out information about the key.
5. The master key used for broadcast encryption [124] are the same which can lead to impersonation attacks.
6. The E0 stream cipher algorithm used for conventional Bluetooth are weak
7. Once the address is captured and associated with a particular user, it can be means of undermining that user's privacy.
8. Legacy pairings provide no passive eavesdropping protection
9. Keys need to be stored securely
10. There is a high degree of freedom for generating the pseudo random numbers which could be weak and follow decipherable patterns
11. Only device authentication is provided and there is no user authentication
12. Secure Bluetooth links are not sufficient, end-to-end security mechanism is needed
13. There is always more to add to Bluetooth security stack such as non-repudiation
14. Application to VANETs requires the nodes to be discoverable all the time which can open the gate for multiple attacks! Usually in conventional application of Bluetooth devices are in the discoverable mode for a short period of time.
15. Overhead is always the big drawback of complicated security mechanism, specifically in VANETs that are more sensitive to overheads and the delays imposed by them. To compromise on overhead as well as the resources, the devices that are already paired do not go through the authentication process. This authenticated device can be used for malicious purposes later on.

The vulnerabilities mentioned above open the way to multifarious attacks such as:

**Bluesnarfing** [125]: In this attack, the adversary uses the device's international mobile equipment identity (IMEI) number to route all the calls from the user's device to the attacker's device. Having access to the storage is very important for running this attack and secure storage techniques could be a proper countermeasure.

**Bluejacking** [126, 127]: Bluejacking is less of a harmful attack resembling fishing in software security. This type of attack is not harmful unless the content sent by the attacker conducts a more complex attack.

**Bluebugging** [128]: This is type of attack is a rather strong attack developed based on a bug on older versions of Bluetooth where the attacker finds access to the commands of the Bluetooth device and can run them independent from the user's input.

**Car Whisperer** [129]: White-hackers used a non-random key in car kits to break into Bluetooth and send audio to or receive it from microphones in the car.

**DoS** [128]: DoS attacks are general less complicated types of attacks that focus on occupying and draining the resources until the device stops functioning properly.

**Fuzzing Attacks** [130]: This is a simpler type of attack that can be performed with less expertise or access to the resources. The attacker simply crafts mal-formed messages and sends it to the device to keep the device busy decoding these non-proper messages.

**SSP** [131]: This type of attack forces the device to switch to Just Works mode. As stated earlier this mode is prone to MITM attacks, hence, impersonation attacks can be performed by the attacker.

**Bluesniff** [117]: a packet sniffing by reverse engineering the hopping pattern of the FH system at a Bluetooth transceiver by extracting the MAC address (MAC address is used for generating the pseudo random hopping pattern in Bluetooth protocol) implemented on Ubertooth.

**Authentication abuse** [122]: In Bluetooth mode 3, since authorization check is not performed, an authenticated remote device can access a Bluetooth service without the device owner's knowledge which is called "authentication abuse".

Table 9 associates each of these attacks to the vulnerabilities of Bluetooth mentioned by NIST.

**Table 9** Cyberattacks/Root cause of cyberattacks on Bluetooth.

| Cyber attack | Vulnerability # | Ref. |
|---|---|---|
| Bluesnarfing | 5,9,10,11 | [125] |
| Bluejacking | 11,13,14 | [126], [127] |
| Bluebugging | 8,11,13,14 | [128] |
| Car Whisperer | 5,7,9,11,15 | [129] |
| DoS | 4,11,13,14,15 | [128] |
| Fuzzing Attacks | 2,6,11,14,15 | [130] |
| SSP Attacks | 1,2,3,6,8,11 | [131] |
| Bluesniff | 9,10,12,14,15 | [117] |
| Authentication abuse | 11,14 | [122] |

Even though Bluetooth wireless technology has a high potential for building a platform for VANETs, there exist numerous security vulnerabilities in the implemented versions of it outside of VANETs. These vulnerabilities and the attacks on a Bluetooth device associated with these vulnerabilities are detailed in this section. Not only the general wireless networking vulnerabilities such as DoS attacks, eavesdropping, MITM attacks, and message modifications, need to be concerned, but also more specific attacks exist that pertain to Bluetooth only which are listed in table 9. Last, even if strong security patches are defined for the wireless connection, the wired connections in the end-to-end system are not protected and a higher layer security solution is required next to a secure Bluetooth and IEEE 802.11 [132, p. 802].

*C. Satellite Radio (2.345-2.60 GHz):*

Satellite radio is originally defined for broadcasting satellite services intended for occupants of vehicles. This, as well as wide coverage on geographical regions that normally fall out of coverage regions of cellular networks makes satellite radio a strong potential for vehicular communications where the vehicles as the main nodes are highly dynamic and cannot be restricted to a specific region. It is not sufficient to use only satellite radio to provide coverage for VANETs however; this technology needs to be augmented with the existing terrestrial access technologies to fill the coverage gap that they bring along. Satellite augmentation is used for enhancement of performance of a standalone GPS system [133] or messages in vehicular networks such as SafeTRIP [134] in S-band, UMTS or directly between vehicles. Here are the applications that satellite augmentation can bring into vehicular communications reported in [134]: (1) communications with the control center, sensor data transmission (temperature/humidity), (2) automatic emergency calls (with video) in case of accident, (3) stolen vehicle tracking, (4) road safety alerts, (5) collaborative road alerts, (6) real-time tracking of vehicles: vehicle tracking, fleet management, (7) live TV/audio, (8) multimedia datacast, (9) driver alertness service (10) information about nearby points of interest.

Satellite radio channels can provide as high bandwidth as 90 MHz however high latency, hundreds of milliseconds, is one of its main drawbacks [135]. This technology avoids the CAPEX concerning the low populated areas, it is ecologic oriented and provides global coverage (high scalability). Large antenna dimension is another potential obstacle standing on the way of using satellite radios for V2I communications. Authors at [136] propose Ku-band geostationary earth orbit (GEO) satellite to encounter the size problem by shifting to slightly high frequencies. The conventional satellite radio uses the 2.3 GHz S-band, while Ku-bands lie between 12 GHz and 18 GHz which helps to solve the interference problem with satellite broadcast which is the main use for satellite radio. Phased array antennas which are strong enough to provide high throughput communication can be designed in Ku-bands and implemented on top of the moving vehicles without drastically changing the aesthetics of the current vehicles [137].

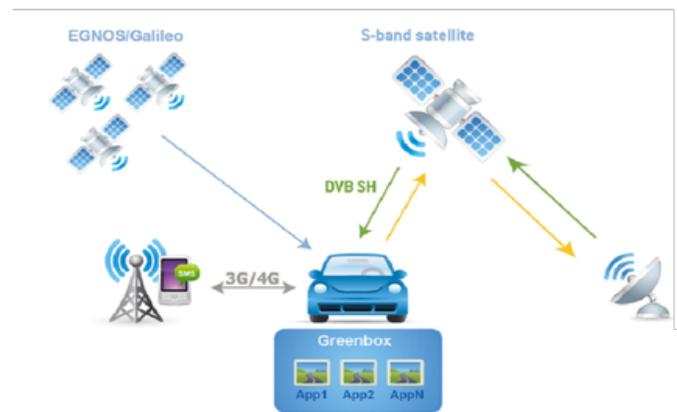

**Figure 6** Satellite augmentation architecture. DVB-SH (Digital Video Broadcasting - Satellite services to Handhelds) is a physical layer standard for delivering IP based media content, based on a hybrid satellite/terrestrial communication [134].

Such hybrid networks with satellite radio as a component have major security issues due to the following [138]. First, the broadcast nature of satellite communication is a security threat. Second, not all security mechanisms are applicable to satellite communications. High latency of these systems requires the security mechanism to impose as low additional latency as possible to the system. Third, satellite messages are prone to burst errors, so the security patch needs to be robust enough to handle burst errors. More details are given on the attacks/vulnerabilities that come along with satellite communications.

***Confidentiality of information*:** the broadcast nature of satellite communications [138] provides the opportunity for an eavesdropper to establish a channel and listen to the information. A performance evaluation of satellite communications in presence of an eavesdropper is given at [139] where sensitivity to ionospheric conditions (common phenomenon at high frequencies) are also taken into account. The latter is a physical layer perspective framework. Physical layer solutions as well as network/transport layer solutions are used to maintain confidentiality of data.

*Spoofing*: Integrity of satellite data can be undertaken by an adversarial party who is able to generate spoofing data. Physical layer/network layer methods are used to combat this type of attack which are explained in detail after covering the attacks in the sequel.

*Session hijacking*: A type of man in the middle attack where an attacker steals a valid user's session token which identifies the user to the server and gives them access to their account. Using malwares is a common tool for a hijacker. Network layer security solutions are given in the sequel to target this type of security breach [140].

*Authentication*: TCP PEP, and SSL implemented together.

*Denial-of-service*: a sufficiently well-equipped adversary can send commands to the satellite and jam or disrupt the communication

To thwart such attacks network layer-transport layer and physical layer solutions have been investigated.

*Network layer-Transport layer solutions:* Internet Security Protocol (IPSec) or Secure Socket Layer (SSL), which are originally designed to provide end-to-end unicast security at transport layer for terrestrial communications, face issues like high delays or malfunctioning of the entire protocol, when applied to the aforementioned hybrid networks.

Performance Enhancing Proxy (PEP) is widely deployed in satellite networks today. It compensates for the large delay caused by slow TCP when operating with satellites with high delay, simply by sending early acknowledgements. IPSec however is not able to coexist with PEP. IPsec has been originally defined for point to point communications. It does not allow for authentication in an intermediate node. Hence TCP PEP which is an intermediate node cannot read or decrypt the TCP header. If used alone, however, IPSec adds 34 bytes of overhead when providing both security and authentication. The same happens to HTTP proxy, and it cannot operate in presence of IPSec. As for the SSL, its conventional use in terrestrial communications is to secure the HTTP connection when needed with encryption, which is called HTTPS. In satellite communications though, SSL encryption does not allow the HTTP proxy to function correctly. TCP PEP can function correctly if only SSL is used.

In order to use the existing network layer security protocols in Satellite communications, modifications need to be made due to the shortcomings mentioned above. A hierarchical approach to key management in hybrid networks is introduced in [138] which extends the state of the art solutions contributing to this problem.

*Physical layer solutions:* The physical layer security measures are usually evaluated using a metric named "secrecy rate" which is an information theoretic measure of the amount of resources that are being used for the legit communication compared to the eavesdropping channel. Deviation from a predefined secrecy rate can be used as an indicator of malicious activities on a channel. On the other hand there are many physical layer solutions that target spoofing attacks on satellite radio [141]. For instance angle of arrival (output of an array of antennas element) is used as a metric for detecting anomalies in the communication [142]. Another example of designing intrusion detection systems for satellite radio signals is given at [143] where fingerprinting based on the variations in the signal levels is used for detection of malicious activities.

### D. Visible Light Communication (VLC) IEEE 802.15.7:

VLC uses the visible light frequencies 430-790 THz as a communication variant. LEDs are mainly used as the intensity modulation sources due to their fast switching time and support up to 500 Mbps for short distances. Recently it has been demonstrated that using similar techniques to MIMO-OFDM of radio frequency the data rate of a VLC link in an indoor scenario can raise up to 1Gb/s [144]. IEEE 802.15.7 provides specifications for achieving a high data rate and flicker free communication [145, p. 7]. Even though the idea of using VLC as the access technology for VANETs is still under consideration and has not been implemented yet, using LEDs in transportation systems dates back to 1998 [146, 147]. Internet connectivity has always been considered as the strongest application for VLC.

In comparison with other access technologies, much higher data rate, low power consumption, low cost, no interference with already crowded radio frequency are the advantages that come with deploying VLC into VANETs. It is originally a NON-LoS communication technology which makes it less prone to multipath distortion [130]. On the other hand, highly contaminated environment with noise and annoyance caused by flickering light, degradations in performance in non LOS communications, are the drawback associated with VLC [148]. As reported by Căilean et al. in [149] the solar noise in the background is 1000 times stronger than the VLC link containing data. In VLC it is not easy to establish bidirectional links between more than two transceivers. Hence, if VLC is to be used for communicating vehicles, each vehicle should be assumed to be a relay node and pass the information to the other vehicles. This is how information is propagated by VLC in VANETs. The latter makes VLC more applicable for broadcasting safety critical messages which are usually in broadcast format and do not require acknowledgement. Specific coding schemes, like Manchester coding which balance the number of 0s and 1s can be used for mitigating the flickering problem which ends up with high bandwidth consumption [149, 150].

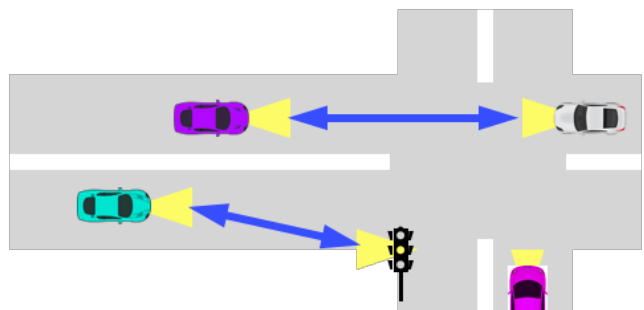

**Figure 7** VLC as a communication medium for VANETs.

VLC shows a huge potential for I2V communications relying on the experiments conducted in [151], [152] where the authors study the channel quality between the street/traffic light and vehicle, and see a stable communication for range of up to 50 meters. For V2I communications [153] reports a link of 90 m with 4.8 Kbps. For V2V communications though, the range of a strong link is demonstrated to be multiple tens of meters based on [133]. Other than having a high potential for vehicular communications, VLC is feasible for platooning [154] and can be a strong candidate to replace GPS [156]. Other researchers have proposed integration of VLC into DSRC so as to take advantage of the benefits that come by each technology [157, 158]. Căilean et al. [149] summarizes the challenges for VLC usage in vehicular applications as: (1) increasing the robustness to noise; (2) increasing the communication range; (3) enhancing mobility; (4) performing distance measurements and visible light positioning; (5) increasing data rate; (6) developing parallel VLC; and (7) developing heterogeneous dedicated short range communications and VLC networks.

Unlike radio frequency based technologies, visible lights cannot penetrate through objects, hence, it can facilitate secure communication in indoor environments [159]. However, just like other access technologies, there exist certain security issues specific to VLC that can affect VANETs.

*Jamming*: The physics of VLC channels are based upon transmission of information via photons, hence, these channels are modeled by Poisson process and their overall throughput does not increase monotonically by the number of users; it has an apex and saturates afterwards. This feature enables the jammers to saturate the channel by adding different sources of light. Other than that, a jamming attack can be simply launched by an adversary capable of achieving a higher illumination at the receiver than the legitimate communication link. Due to high directionality of visible light channels, it is not easy for the neighboring nodes to detect the presence of a jammer, hence, in large scale deployments, it is not possible to detect a jamming point in the network and propagate its location so as to avoid it by simply rerouting the traffic [160].

*Snooping and data modification*: The naive assumption that due to directionality and visibility, snooping and data modification are difficult to do, does not mean they are impossible. Most of the MAC layer protection is assumed to be provided via cryptography which is prone to a wide range of attacks, not to mention the fact that in most cases manufacturers skip the cryptography at implementation or leave it to an optional mode based on user preference. The latter mostly stems from the lack of computational power. The security mechanisms suggested at IEEE 802.15.7 provide authenticity and confidentiality, but do not provide protection mechanisms for non-repudiation and integrity. Authors at [161] perform simulations as well as experiments in order to prove that data sniffing is feasible over an open VLC channel.

*Masquerade attacks*: The current version of standard provides protection against the outsider adversaries and there is no security check for a malicious user within the group that share the peer-to-peer communication key. This opens up opportunities for an internal source to be compromised and further used for malicious purposes via what is so called masquerade attacks.

*Side-channel attacks*: Since the data is modulated into the intensity level of the emitted light, the power consumption of the electric device changes to high and low accordingly. This power consumption pattern can be a means of intrusion for an adversary with proper hardware to track and decipher the bits [162].

*Eavesdropping*: Experiments are performed in [163] to demonstrate the ease of eavesdropping without advanced hardware requirement in different scenarios of indoor/outdoor. There exist three well-known security mechanisms that can be used to protect VLC: proximity-based protection, steganographic protection, and cryptographic protection [164].

To prevent such attacks proximity based, Steganographic protection, chaffing and winnowing, and cryptographic solutions have been investigated.

*Proximity based solutions:* Such a solution is limited to protected environments that offer snoop-free line-of-sight communication which is a challenging requirement in practical implementations.

*Steganographic Protection, chaffing and winnowing:* In this type of protection information is secretly hidden inside the existing illumination. Unlike cryptographic solutions, steganography is more about confidentiality than integrity or authentication and it is less expensive to implement. Worth noting that violation of confidentiality directly relates to violation of integrity as well. Chaffing and winnowing is another method that can be used to provide authentication and confidentiality without using cryptography [165]. These methods are strong candidates to provide security when the resources to implement strong cryptos are not available.

*Cryptographic solutions:* IEEE 802.15.7 proposes to place the cryptographic solutions on MAC layer to provide integrity and authentication. The research in the field of cryptography is usually narrowed down to secret key generation schemes which happen to be difficult to generate and vulnerable to vast variety of attacks. In ultrawideband channels however, researchers have proposed to exploit the channel characteristics to generate this key in a secure manner at [166]. To be more specific, due to channel reciprocity both the transmitter and the receiver experience the same channel propagation effects such as shadowing, fading or path loss, and this can be transformed into a metric and used for key generation in a non-mimicable way. VLC however, is less prone to multipath effect compared to radio frequency, hence, the dominant propagation effect in VLC is pathloss, which might be easier for an adversarial party to replicate one they know the characteristics of the environment.

### E. LTE/5G:

Currently researchers have attended to investigate the challenges in regard to the 5G networks (e.g., [167]). However, the focus of the survey is only limited to the 5G performance challenges rather the security aspects of it. The future of communication systems is entangled with 5G. As 5G promises ultralow latency, ultrahigh capacity, infinite bandwidth, etc., the security challenges will be dramatic [168]. Heterogeneity of the technologies as well as the devices involved in 5G is the main reason for this, which might lead to security issues that are not recognized yet. It is of utmost importance to shed light on the security vulnerabilities of this technology and take them into account by design from inception, because 5G will have to tackle with much stronger threats. Mantas et al. [169] categorizes the main targets of attackers as: users or nodes, access networks, core networks and external IP networks.

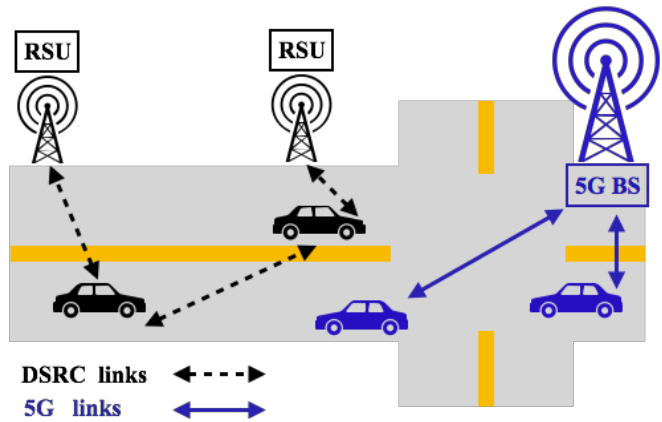

**Figure 8** DSRC vs. LTE/5G for VANETs.

Open operating systems in 5G will allow nodes to install applications from all adversarial/benign parties without having a well-defined infrastructure for authentication, in addition to the denial of service attacks which are very common on end users. Vehicles as the main nodes in vehicular communications, are vulnerable to numerous attacks via infotainment system/Bluetooth which target their internal network (such as CAN, FlexRay, etx.) which are elaborated in our previous work [9]. Malwares can undertake privacy of the user by accessing information such as location of the vehicle. Botnets on the other hand will be extensively used by attackers to take control of the vehicles due to openness of 5G to different connectivity options in uplink [170]. The compromised vehicles can be used later by the attackers as a new foothold to launch DoS attacks or congestion attacks on the other vehicles in the network [171].

The focus of this work however is on how the access technologies that come with 5G will bring in new security challenges. Heterogeneity of access technologies, creates possibility for interoperability of devices equipped with 4G, 3G and 2G, this can open the way for an attacker to switch to 2G or 3G by announcing unavailability of 4G and shifting the communication to 2G or 3G, hence taking advantage of the security vulnerabilities that come along the older access technologies. 4G as well, brings lots of security issues which are summarized in the table below.

The radio resource control (RRC) and the handover confirm messages in 4G are not encrypted. This can help an attacker to access Cell Radio Network Temporary Identifier (C-RNTI) for each user and do location tracking [172]. Location tracking can also be done via following the packet sequence numbers and tracking them since they are continuous numbers [173].

False buffer status report attacks are a subset of denial of service attacks which targets the load balancing, scheduling and admission control algorithms by sending false buffer scheduling messages and tricking the algorithm to allocate more resources to the attacker [172, 173]. The proposed solution for this problem is to use token based unique identifiers to provide authentication at MAC layer buffer status messages.

Discontinuous Reception Period (DRX) is a period where a node can be in active mode however turn off the transceiver for reduced power consumption. This can be a foothold for a malicious node to insert messages. Table 10 provides a summary of these attacks and their corresponding footholds. Interested reader might see [172] for more information on MAC layer vulnerabilities of 4G standards (WiMAX and LTE).

**Table 10** Footholds of cyberattacks on LTE/5G.

| Attack | Foothold | Ref. |
|---|---|---|
| Location tracking | Cell Radio Network Temporary Identifier (C-RNTI); Packet Sequence Numbers | [172] [173] |
| False buffer status report attacks | Lack of authentication in MAC layer false buffer messages | [172] [173] |
| Message insertion attacks | Discontinuous Reception Period | [172] [173] |

## V. COMPARISON

Making a solid decision on which access technology to choose is a non-trivial task. The tables below summarize the specifications of each technology and shed light on the security issues that come along with them. Table 12 more specifically gives a summary of the pros and cons of using each access technology for VANETs. Depending on the application, sometimes more than one technology should be used in an augmented solution, so as to benefit from the diverse advantages that they have. These are called hybrid solutions. The satellite radio for instance, is a strong candidate for being augmented with DSRC or 5G for a global coverage of communication links. VLC can provide high data rates with a low cost infrastructure for V2I communications in combination with any other technology to support V2V long range links.

## VI. CONCLUSION

VANET is projected to augment safety, comfort, transportation efficiency, and to overcome the environmental impacts of transportation, but at the presence of security footholds all the advantages can dim. In the presence of any weaknesses, attackers can exploit VANETs' availability, authentication, identification, confidentiality, integrity, data trust, privacy, and non-repudiation security goals. A successful security protocol shall be compliant with VANET security requirements with an all-inclusive approach. Moreover, there are many questions that need to be resolved prior to VANET's large-scale implementation. For instance: (1) What would be the impact of known/unknown cyberattacks on transportation? (2) How a road user should react to the malfunctioned behavior of the car? (3) Do road users need to be trained for cyber intrusion events? and (4) Who should road users call for fixing their cars due to a cyberattack? Security is a core technical challenge of the VANETs that if not appropriately executed, will drastically affect ITS network deployment. Prior to a real-world deployment, security concerns should be raised and addressed throughout the ITS domain, or dire consequences will be inevitable [9, 10]. This should not only cover the security issues that exist in a general ad hoc network, but also cover the specific security issues that come along based on the access technology that is used in the VANET. Lessons learned from this study depict that road users' safety, privacy and security can be compromised due to the VANET vulnerabilities. Adversaries not only can take control of vehicles but can compromise the traffic safety and leave a driver in a traumatic state. Eventually adversaries will supersede drivers to remotely-drive vehicles in the near future. We argue that due to the current gap in the literature with a focus on evaluating attacks' consequences, security attack impacts on transportation network shall be explored via stated and revealed methods (e.g., [174]–[179]).

Table 11 Comparison of the available access technologies for VANETs.

| Access Technology | Coverage | Data Rate | BW | Status | General Security Issues | Specific Security Issues |
|---|---|---|---|---|---|---|
| DSRC | 1000 m | 27 Mbps | 75 MHz | Incomplete | Availability (Table 2) Authentication (Tables 3) Privacy (Table 3) Confidentiality (Table 4) Non-repudiation (Table 5) | See Table 6 |
| Bluetooth | 10-100 m | 24 Mbps | 78 MHz | Implemented | Eavesdropping MITM Impersonation Secure storage Confidentiality (Table 4) Authentication (Table 3) Non-repudiation (Table 5) Integrity (Table 4) | See Table 9 |
| Satellite | 35e6 m | 4 to 229 kilobits-per-second; Changes with channel | 90 MHz | Implemented | Confidentiality (Table 4) Authentication (Table 3) DoS MITM | Spoofing; Session Hijacking |
| VLC | 50 m (V2I); 20 m (V2V) | 500-1000 Mbps | 360 THz | Not implemented, Available | Eavesdropping Confidentiality (Table 4) Authentication (Table 3) Non-repudiation (Table 5) Integrity (Table 4) | Snooping Masquerade Side channel attack Jamming |
| LTE/5G | 200-300 m | 10 Gbps | 270 GHz | Not implemented, Available | DoS; Congestion attacks | See Table 10 |

Table 12 Pros and Cons of the available access technologies

| Access Technology | Pros | Cons |
|---|---|---|
| DSRC | Wide Coverage | See Table 6 |
| Bluetooth | Good for dense areas; High power efficiency | Latency |
| Satellite | Low cost; Extremely wide coverage (even the blind zones) | Hundreds of milliseconds of delay; Burst errors |
| VLC | High data rate Low cost; No interference with radio frequency; Low power consumption | High contaminated environment to noise Restricted field of view; Sensitive to flickering light; More suitable for broadcasting scenarios; Unable to provide bidirectional channels |
| LTE/5G | Low latency of 5 ms (V2I); Low latency of 100 ms (V2V); High Coverage | The mmWave communication needs LOS |

This study sheds light on the new security issues that appear by application of different access technologies after fully covering the security challenges and their corresponding countermeasures from a general ad hoc network point of view. We contributed to the current literature of the CAV security by incorporating VANET and its access technology security challenges and comparing the different access technologies performance, security challenges and propound heterogeneous technologies. The access technologies that are covered here are DSRC, Bluetooth, Satellite radio, VLC, and 5G, however, new technologies can be proposed later since VANETs are at their infancy. There are multiple pros and cons pertaining to each technology and there is no solid formula for deciding on which technology is the best. The latter highly depends on application, demand, scale of the network and even geographical location, since VANETs are highly dynamic and diverse and cannot be limited to specific frame of a certain location or application.

The types of security challenges in VANETs change extensively depending on the types of technologies being used. For instance, in a dense and low power scenario Bluetooth can be a good candidate for VANETs, however it is vulnerable to numerous attacks and its security issues need to be considered carefully. On the other hand, Bluetooth is already suffering from high latency and the security mechanisms associated with it need to be chosen carefully with a focus on minimizing the additional overhead. Another example is highways and roads that are located far away from cities and consequently inaccessible. Satellite radio is a strong candidate to provide coverage for such blind zones. However, it is not fully compatible with the security tools that have already been implemented and research needs to be done to fill in this security gap. One way to find a solution which meets the demands of VANETs is to augment technologies so as to take advantage of the different capabilities that they bring along. In this case, security mechanisms need to be combined together with ultimate elaboration and intricacy which requires significant effort.

Not only the industrial parties and researchers, but also decision makers and policy makers need to be aware of the abundant security issues when VANETs are concerned. Unfortunately, the main focus so far has been on the performance and security has been overlooked. The security challenges that come with VANETs are either neglected or naively limited to general ad hoc network security issues that threaten VANETs. The main goal of this document is to provide a substantial framework which covers the security issues for all types of VANETs and provide the reader with necessary technical details when the access technology changes in these networks.

VII. REFERENCE

[1] "Hackers Remotely Kill a Jeep on the Highway—With Me in It," *WIRED*. [Online]. Available: https://www.wired.com/2015/07/hackers-remotely-kill-jeep-highway/. [Accessed: 12-Oct-2017].

[2] Satsafe, "How Hughes Telematics Works," *Satsafe*, 17-Mar-2017. .

[3] "The Dangers of the Hackable Car - WSJ." [Online]. Available: https://www-wsj-com.cdn.ampproject.org/c/s/www.wsj.com/amp/articles/the-dangers-of-the-hackable-car-1505700481. [Accessed: 18-Oct-2017].

[4] H. Kawashima, "Japanese perspective of driver information systems," *Transportation*, vol. 17, no. 3, pp. 263–284, 1990.

[5] H. Hartenstein and K. P. Laberteaux, "A tutorial survey on vehicular ad hoc networks," *Communications Magazine, IEEE*, vol. 46, no. 6, pp. 164–171, 2008.

[6] A.-M. Cailean, B. Cagneau, L. Chassagne, V. Popa, and M. Dimian, "A survey on the usage of DSRC and VLC in communication-based vehicle safety applications," in *Communications and Vehicular Technology in the Benelux (SCVT), 2014 IEEE 21st Symposium on*, 2014, pp. 69–74.

[7] M. Raya, P. Papadimitratos, and J.-P. Hubaux, "Securing vehicular communications," *IEEE Wireless Communications Magazine, Special Issue on Inter-Vehicular Communications*, vol. 13, no. LCA-ARTICLE-2006-015, pp. 8–15, 2006.

[8] J. Blum and A. Eskandarian, "The threat of intelligent collisions," *IT professional*, vol. 6, no. 1, pp. 24–29, 2004.

[9] K. B. Kelarestaghi, M. Foruhandeh, K. Heaslip, and R. Gerdes, "Intelligent Transportation System Security: Impact-Oriented Risk Assessment of In-Vehicle Networks," IEEE Intelligent Transportation Systems Magazine, doi: 10.1109/MITS.2018.2889714, 2019.

[10] K. B. Kelarestaghi, K. Heaslip, V. Fessmann, M. Khalilikhah, and A. Fuentes, "Intelligent transportation system security: hacked message signs," SAE International Journal of Transportation Cybersecurity & Privacy, vol. 1, no. 2, doi:10.4271/11-01-02-0004, 2018.

[11] K. B. Kelarestaghi, W. Zhang, Y. Wang, L. Xiao, K. Hancock, and K. P. Heaslip, "Impacts to crash severity outcome due to adverse weather and other causation factors," *Advances in transportation studies*, vol. 43, pp. 31–42, 2017.

[12] W. Zhang, L. Xiao, Y. Wang, and K. Kelarestaghi, "Big Data Approach of Crash Prediction," presented at the Transportation Research Board 97th Annual Meeting, no. 18-04956, 2018.

[13] S. Banerjee and K. N. Khadem, "Factors Influencing Injury Severity in Alcohol-Related Crashes: A Neural Network Approach Using HSIS Crash Data," *ITE Journal*, vol. 2019, p. 89.

[14] M. R. Islam, K. B. Kelarestaghi, A. Ermagun, and S. Banerjee, "Gender Differences in Injury Severity Risk of Single-Vehicle Crashes in Virginia: A Nested Logit Analysis of Heterogeneity," *arXiv preprint arXiv:1901.03289*, 2019.

[15] F. C. Commission and others, "FCC Allocates Spectrum in 5.9 GHz Range for Intelligent Transportation Systems Uses," *Report No. ET*, pp. 99–5, 1999.

[16] S. Al-Sultan, M. M. Al-Doori, A. H. Al-Bayatti, and H. Zedan, "A comprehensive survey on vehicular Ad Hoc network," *Journal of network and computer applications*, vol. 37, pp. 380–392, 2014.

[17] V. Yadav, S. Misra, and M. Afaque, "Security in vehicular ad hoc networks," *Security of Self-Organizing Networks: MANET, WSN, WMN, VANET*, p. 227, 2010.

[18] N. Gupta, A. Prakash, and R. Tripathi, "Medium access control protocols for safety applications in Vehicular Ad-Hoc Network: A classification and comprehensive survey," *Vehicular Communications*, vol. 2, no. 4, pp. 223–237, 2015.

[19] A. Indra and R. Murali, "Routing protocols for vehicular adhoc networks (VANETs): A review," 2014.

[20] A. Dhamgaye and N. Chavhan, "Survey on security challenges in VANET," *International Journal of Computer Science and Network*, vol. 2, no. 1, 2013.

[21] S. Zeadally, R. Hunt, Y.-S. Chen, A. Irwin, and A. Hassan, "Vehicular ad hoc networks (VANETS): status, results, and challenges," *Telecommunication Systems*, vol. 50, no. 4, pp. 217–241, 2012.

[22] M. N. Mejri, J. Ben-Othman, and M. Hamdi, "Survey on VANET security challenges and possible cryptographic solutions," *Vehicular Communications*, vol. 1, no. 2, pp. 53–66, 2014.

[23] J. M. de Fuentes, A. I. González-Tablas, and A. Ribagorda, "Overview of security issues in Vehicular Ad-hoc Networks," 2010.

[24] H. Moustafa and Y. Zhang, *Vehicular networks: techniques, standards, and applications*. Auerbach publications, 2009.

[25] J. Jakubiak and Y. Koucheryavy, "State of the art and research challenges for VANETs," in *Consumer communications and networking conference, 2008. CCNC 2008. 5th IEEE*, 2008, pp. 912–916.

[26] L. Wischhof, A. Ebner, and H. Rohling, "Information dissemination in self-organizing intervehicle networks," *Intelligent Transportation Systems, IEEE Transactions on*, vol. 6, no. 1, pp. 90–101, 2005.

[27] Y. Toor, P. Muhlethaler, and A. Laouiti, "Vehicle ad hoc networks: applications and related technical issues," *Communications Surveys & Tutorials, IEEE*, vol. 10, no. 3, pp. 74–88, 2008.

[28] K. Timpone, J. Walker, and K. Dopart, "Connected vehicle applications: safety," US Department of Transportation, FHWA-JPO-16-427, Feb. 2017.

[29] US Department of Transportation, "Connected Vehicle Reference Implementation Architecture," 27-Feb-2017. [Online]. Available: http://local.iteris.com/cvria/html/applications/applications.html.

[30] B. Parno and A. Perrig, "Challenges in securing vehicular networks," in *Workshop on hot topics in networks (HotNets-IV)*, 2005, pp. 1–6.

[31] M. Raya and J.-P. Hubaux, "Securing vehicular ad hoc networks," *Journal of computer security*, vol. 15, no. 1, pp. 39–68, 2007.

[32] M. Zhao, J. Walker, and C.-C. Wang, "Challenges and Opportunities for Securing Intelligent Transportation System," *Emerging and Selected Topics in Circuits and Systems, IEEE Journal on*, vol. 3, no. 1, pp. 96–105, 2013.

[33] T. Foss, "Safe and secure Intelligent Transport Systems (ITS)," in *Transport Research Arena (TRA) 5th Conference: Transport Solutions from Research to Deployment*, 2014.

[34] M. Foruhandeh and R. Gerdes, "Analysis of authentication parameters in uplink scenario of Heterogeneous Networks," in *2017 IEEE Trustcom/BigDataSE/ICESS*, 2017, pp. 833–838.

[35] M. Foruhandeh, N. Tadayon, and S. Aïssa, "Uplink Modeling of K-Tier Heterogeneous Networks: A Queuing Theory Approach," *IEEE Communications Letters*, vol. 21, no. 1, pp. 164–167, DOI: 10.1109/LCOMM.2016.2619338, 2017.

[36] A. Aijaz *et al.*, "Attacks on inter vehicle communication systems-an analysis," in *Proc. International Workshop on Intelligent Transportation (WIT'2006), Hamburg, Germany*, 2006, pp. 189–194.

[37] I. A. Sumra, H. B. Hasbullah, and J. B. AbManan, "Attacks on Security Goals (Confidentiality, Integrity, Availability) in VANET: A Survey," in *Vehicular Ad-hoc Networks for Smart Cities*, Springer, 2015, pp. 51–61.

[38] N. Ekedebe, W. Yu, H. Song, and C. Lu, "On a simulation study of cyber attacks on vehicle-to-infrastructure communication (V2I) in Intelligent Transportation System (ITS)," in *SPIE Sensing Technology+ Applications*, 2015, pp. 94970B–94970B.

[39] M. Whaiduzzaman, M. Sookhak, A. Gani, and R. Buyya, "A survey on vehicular cloud computing," *Journal of Network and Computer Applications*, vol. 40, pp. 325–344, 2014.

[40] M. E. Mathew and P. ARK, "Threat analysis and defence mechanisms in VANET," *Int. J. Adv. Res. Comput. Sci. Softw. Eng*, vol. 3, no. 1, pp. 47–53, 2013.

[41] X. Lin, X. Sun, P.-H. Ho, and X. Shen, "GSIS: a secure and privacy-preserving protocol for vehicular communications," *Vehicular Technology, IEEE Transactions on*, vol. 56, no. 6, pp. 3442–3456, 2007.

[42] Y.-C. Hu, A. Perrig, and D. B. Johnson, "Wormhole attacks in wireless networks," *IEEE Journal on Selected Areas in Communications*, vol. 24, no. 2, pp. 370–380, Feb. 2006.

[43] L. He and W. T. Zhu, "Mitigating DoS attacks against signature-based authentication in VANETs," in *Computer Science and Automation Engineering (CSAE), 2012 IEEE International Conference on*, 2012, vol. 3, pp. 261–265.

[44] S. Biswas, J. Mišic, and V. B. Misic, "DDoS attack on WAVE-enabled VANET through synchronization," in *Global Communications Conference (GLOBECOM), 2012 IEEE*, 2012, pp. 1079–1084.

[45] J. Domingo-Ferrer and Q. Wu, "Safety and privacy in vehicular communications," in *Privacy in Location-Based Applications*, Springer, 2009, pp. 173–189.

[46] S. Xia and J. You, "A group signature scheme with strong separability," *Journal of Systems and Software*, vol. 60, no. 3, pp. 177–182, 2002.

[47] V. Daza, J. Domingo-Ferrer, F. Sebé, and A. Viejo, "Trustworthy privacy-preserving car-generated announcements in vehicular ad hoc networks," *Vehicular Technology, IEEE Transactions on*, vol. 58, no. 4, pp. 1876–1886, 2009.

[48] M. S. Al-kahtani, "Survey on security attacks in Vehicular Ad hoc Networks (VANETs)," in *Signal Processing and Communication Systems (ICSPCS), 2012 6th International Conference on*, 2012, pp. 1–9.

[49] Y.-C. Hu, A. Perrig, and D. B. Johnson, "Packet leashes: a defense against wormhole attacks in wireless networks," in *INFOCOM 2003. Twenty-Second Annual Joint Conference of the IEEE Computer and Communications. IEEE Societies*, 2003, vol. 3, pp. 1976–1986.

[50] B. Cherkaoui, A. Beni-hssane, and M. Erritali, "A Clustering Algorithm for Detecting and Handling Black Hole Attack in Vehicular Ad Hoc Networks," in *Europe and MENA Cooperation Advances in Information and Communication Technologies*, Springer, 2017, pp. 481–490.

[51] M. Nogueira, H. Silva, A. Santos, and G. Pujolle, "A security management architecture for supporting routing services on WANETs," *IEEE Transactions on Network and Service Management*, vol. 9, no. 2, pp. 156–168, 2012.

[52] S. Verma, B. Mallick, and P. Verma, "Impact of gray hole attack in VANET," in *Next Generation Computing Technologies (NGCT), 2015 1st International Conference on*, 2015, pp. 127–130.

[53] A. Hamieh, J. Ben-Othman, A. Gueroui, and F. Naït-Abdesselam, "Detecting greedy behaviors by linear regression in wireless ad hoc networks," in *Communications, 2009. ICC'09. IEEE International Conference on*, 2009, pp. 1–6.

[54] A. Dhamgaye and N. Chavhan, "Survey on security challenges in VANET 1," 2013.

[55] R. Minhas and M. Tilal, "Effects of jamming on IEEE 802.11 p systems," 2010.

[56] J.-P. Hubaux, S. Capkun, and J. Luo, "The security and privacy of smart vehicles," *IEEE Security & Privacy*, no. 3, pp. 49–55, 2004.

[57] L. Gollan, I. L. Gollan, and C. Meinel, "Digital signatures for automobiles?!," in *in Systemics, Cybernetics and Informatics (SCI)*, 2002.

[58] J. S. Warner and R. G. Johnston, "GPS spoofing countermeasures," *Homeland Security Journal*, vol. 25, no. 2, pp. 19–27, 2003.

[59] I. A. Sumra, H. B. Hasbullah, and J. B. AbManan, "Attacks on Security Goals (Confidentiality, Integrity, Availability) in VANET: A Survey," in *Vehicular Ad-*

hoc Networks for Smart Cities, Springer, 2015, pp. 51–61.

[60] I. A. Sumra, H. Bin Hasbullah, and J. Bin AbManan, "Effects of attackers and attacks on availability requirement in vehicular network: a survey," in *Computer and Information Sciences (ICCOINS), 2014 International Conference on*, 2014, pp. 1–6.

[61] S. RoselinMary, M. Maheshwari, and M. Thamaraiselvan, "Early detection of dos attacks in VANET using attacked packet detection algorithm (apda)," in *Information Communication and Embedded Systems (ICICES), 2013 International Conference on*, 2013, pp. 237–240.

[62] R. Minhas and M. Tilal, "Effects of jamming on IEEE 802.11 p systems," 2010.

[63] A. Hamieh, J. Ben-Othman, A. Gueroui, and F. Naït-Abdesselam, "Detecting greedy behaviors by linear regression in wireless ad hoc networks," in *Communications, 2009. ICC'09. IEEE International Conference on*, 2009, pp. 1–6.

[64] S. M. Safi, A. Movaghar, and M. Mohammadizadeh, "A novel approach for avoiding wormhole attacks in VANET," in *Internet, 2009. AH-ICI 2009. First Asian Himalayas International Conference on*, 2009, pp. 1–6.

[65] W. R. Pires, T. H. de Paula Figueiredo, H. C. Wong, A. Loureiro, and others, "Malicious node detection in wireless sensor networks," in *Parallel and Distributed Processing Symposium, 2004. Proceedings. 18th International*, 2004, p. 24.

[66] H. Hasrouny, A. E. Samhat, C. Bassil, and A. Laouiti, "VANet security challenges and solutions: A survey," *Vehicular Communications*, 2017.

[67] I. A. Sumra, J.-L. Ab Manan, and H. Hasbullah, "Timing attack in vehicular network," in *Proceedings of the 15th WSEAS International Conference on Computers, World Scientific and Engineering Academy and Society (WSEAS)*, 2011, pp. 151–155.

[68] J. Blum and A. Eskandarian, "The threat of intelligent collisions," *IT professional*, vol. 6, no. 1, pp. 24–29, 2004.

[69] A. Hesham, A. Abdel-Hamid, and M. A. El-Nasr, "A dynamic key distribution protocol for PKI-based VANETs," in *Wireless Days (WD), 2011 IFIP*, 2011, pp. 1–3.

[70] A.-S. K. Pathan, *Security of self-organizing networks: MANET, WSN, WMN, VANET*. CRC press, 2010.

[71] S. Brands and D. Chaum, "Distance-bounding protocols," in *Advances in Cryptology—EUROCRYPT'93*, 1994, pp. 344–359.

[72] N.-W. Lo and H.-C. Tsai, "Illusion attack on vanet applications-a message plausibility problem," in *Globecom Workshops, 2007 IEEE*, 2007, pp. 1–8.

[73] K. Sampigethaya, M. Li, L. Huang, and R. Poovendran, "AMOEBA: Robust location privacy scheme for VANET," *Selected Areas in Communications, IEEE Journal on*, vol. 25, no. 8, pp. 1569–1589, 2007.

[74] E. B. Hamida, H. Noura, and W. Znaidi, "Security of cooperative intelligent transport systems: Standards, threats analysis and cryptographic countermeasures," *Electronics*, vol. 4, no. 3, pp. 380–423, 2015.

[75] S. S. Kaushik, "Review of different approaches for privacy scheme in VANETs," *International Journal*, vol. 5, 2013.

[76] R. Lu, X. Lin, T. H. Luan, X. Liang, and X. Shen, "Pseudonym changing at social spots: An effective strategy for location privacy in vanets," *IEEE transactions on vehicular technology*, vol. 61, no. 1, p. 86, 2012.

[77] J. Freudiger, M. Raya, M. Félegyházi, P. Papadimitratos, and others, "Mix-zones for location privacy in vehicular networks," 2007.

[78] J. T. Isaac, S. Zeadally, and J. S. Camara, "Security attacks and solutions for vehicular ad hoc networks," *Communications, IET*, vol. 4, no. 7, pp. 894–903, 2010.

[79] I. A. Sumra, I. Ahmad, H. Hasbullah, and J. Bin Ab Manan, "Behavior of attacker and some new possible attacks in Vehicular Ad hoc Network (VANET)," in *Ultra Modern Telecommunications and Control Systems and Workshops (ICUMT), 2011 3rd International Congress on*, 2011, pp. 1–8.

[80] J. R. Douceur, "The sybil attack," in *Peer-to-peer Systems*, Springer, 2002, pp. 251–260.

[81] A. Rawat, S. Sharma, and R. Sushil, "VANET: security attacks and its possible solutions," *Journal of Information and Operations Management*, vol. 3, no. 1, pp. 301–304, 2012.

[82] D. Singelee and B. Preneel, "Location verification using secure distance bounding protocols," in *Mobile Adhoc and Sensor Systems Conference, 2005. IEEE International Conference on*, 2005, pp. 7–pp.

[83] B. Xiao, B. Yu, and C. Gao, "Detection and localization of sybil nodes in VANETs," in *Proceedings of the 2006 workshop on Dependability issues in wireless ad hoc networks and sensor networks*, 2006, pp. 1–8.

[84] C. Zhang, X. Lin, R. Lu, and P.-H. Ho, "RAISE: an efficient RSU-aided message authentication scheme in vehicular communication networks," in *Communications, 2008. ICC'08. IEEE International Conference on*, 2008, pp. 1451–1457.

[85] X. Lin, X. Sun, X. Wang, C. Zhang, P.-H. Ho, and X. S. Shen, "TSVC: timed efficient and secure vehicular communications with privacy preserving," *Wireless Communications, IEEE Transactions on*, vol. 7, no. 12, pp. 4987–4998, 2008.

[86] G. Calandriello, P. Papadimitratos, J.-P. Hubaux, and A. Lioy, "Efficient and robust pseudonymous authentication in VANET," in *Proceedings of the fourth ACM international workshop on Vehicular ad hoc networks*, 2007, pp. 19–28.

[87] C. Zhang, R. Lu, X. Lin, P.-H. Ho, and X. Shen, "An efficient identity-based batch verification scheme for vehicular sensor networks," in *INFOCOM 2008. The 27th Conference on Computer Communications. IEEE*, 2008.

[88] A. Studer, F. Bai, B. Bellur, and A. Perrig, "Flexible, extensible, and efficient VANET authentication," *Communications and Networks, Journal of*, vol. 11, no. 6, pp. 574–588, 2009.

[89] A. Wasef and X. Shen, "Efficient group signature scheme supporting batch verification for securing

vehicular networks," in *Communications (ICC), 2010 IEEE International Conference on*, 2010, pp. 1–5.

[90] A. Perrig, R. Canetti, J. D. Tygar, and D. Song, "The TESLA broadcast authentication protocol," *Rsa Cryptobytes*, vol. 5, 2005.

[91] N. O. Tippenhauer, C. Pöpper, K. B. Rasmussen, and S. Capkun, "On the requirements for successful GPS spoofing attacks," in *Proceedings of the 18th ACM conference on Computer and communications security*, 2011, pp. 75–86.

[92] M. Raya, P. Papadimitratos, and J.-P. Hubaux, "Securing vehicular communications," *IEEE wireless communications*, vol. 13, no. 5, 2006.

[93] B. Hoh, M. Gruteser, H. Xiong, and A. Alrabady, "Preserving privacy in gps traces via uncertainty-aware path cloaking," in *Proceedings of the 14th ACM conference on Computer and communications security*, 2007, pp. 161–171.

[94] S. Duri *et al.*, "Framework for security and privacy in automotive telematics," in *Proceedings of the 2nd international workshop on Mobile commerce*, 2002, pp. 25–32.

[95] L. Huang, K. Matsuura, H. Yamane, and K. Sezaki, "Enhancing wireless location privacy using silent period," in *Wireless Communications and Networking Conference, 2005 IEEE*, 2005, vol. 2, pp. 1187–1192.

[96] A. Pathre, "Identification of malicious vehicle in vanet environment from ddos attack," *Journal of Global Research in Computer Science*, vol. 4, no. 6, pp. 30–34, 2013.

[97] C. Langley, R. Lucas, and F. Huirong, "Key management in vehicular ad-hoc networks," in *2008 IEEE International Conference on Electro/Information Technology*, 2008, pp. 223–226.

[98] M. Verma and D. Huang, "SeGCom: secure group communication in VANETs," in *Consumer Communications and Networking Conference, 2009. CCNC 2009. 6th IEEE*, 2009, pp. 1–5.

[99] X. Hong, D. Huang, M. Gerla, and Z. Cao, "SAT: situation-aware trust architecture for vehicular networks," in *Proceedings of the 3rd international workshop on Mobility in the evolving internet architecture*, 2008, pp. 31–36.

[100] J. M. De Fuentes, L. González-Manzano, A. I. González-Tablas, and J. Blasco, "Security models in vehicular ad-hoc networks: A survey," *IETE Technical Review*, vol. 31, no. 1, pp. 47–64, 2014.

[101] S. Park and C. C. Zou, "Reliable traffic information propagation in vehicular ad-hoc networks," in *Sarnoff Symposium, 2008 IEEE*, 2008, pp. 1–6.

[102] P. Cencioni and R. Di Pietro, "A mechanism to enforce privacy in vehicle-to-infrastructure communication," *Computer communications*, vol. 31, no. 12, pp. 2790–2802, 2008.

[103] M. Raya, A. Aziz, and J.-P. Hubaux, "Efficient secure aggregation in VANETs," in *Proceedings of the 3rd international workshop on Vehicular ad hoc networks*, 2006, pp. 67–75.

[104] H. Harney, "Group key management protocol (GKMP) architecture," *Group*, 1997.

[105] R. G. Engoulou, M. Bellaïche, S. Pierre, and A. Quintero, "VANET security surveys," *Computer Communications*, vol. 44, pp. 1–13, 2014.

[106] K. Plossl, T. Nowey, and C. Mletzko, "Towards a security architecture for vehicular ad hoc networks," in *Availability, Reliability and Security, 2006. ARES 2006. The First International Conference on*, 2006, pp. 8–pp.

[107] M. N. Mejri, J. Ben-Othman, and M. Hamdi, "Survey on VANET security challenges and possible cryptographic solutions," *Vehicular Communications*, vol. 1, no. 2, pp. 53–66, 2014.

[108] IEEE Std 1609.2[TM], "IEEE Standard for Wireless Access in Vehicular Environments—Security Services for Applications and Management Messages." IEEE, 2013.

[109] D. Boneh and H. Shacham, "Group signatures with verifier-local revocation," in *Proceedings of the 11th ACM conference on Computer and communications security*, 2004, pp. 168–177.

[110] K. Sampigethaya, L. Huang, M. Li, R. Poovendran, K. Matsuura, and K. Sezaki, "CARAVAN: Providing location privacy for VANET," DTIC Document, 2005.

[111] R. Baldessari *et al.*, "Car-2-car communication consortium-manifesto," 2007.

[112] S. Olariu and M. C. Weigle, *Vehicular networks: from theory to practice*. Crc Press, 2009.

[113] "802.15.5-2009 - IEEE Recommended Practice for Information technology-- Telecommunications and information exchange between systems-- Local and metropolitan area networks-- Specific requirements Part 15.5: Mesh Topology Capability in Wireless Personal Area Networks (WPANs) - IEEE Standard." [Online]. Available: https://ieeexplore.ieee.org/document/4922106. [Accessed: 16-Dec-2018].

[114] Q. Yang and L. Huang, *Inside Radio: An Attack and Defense Guide*. Springer, 2018.

[115] R. Frank, W. Bronzi, G. Castignani, and T. Engel, "Bluetooth Low Energy: An alternative technology for VANET applications," in *2014 11th Annual Conference on Wireless On-demand Network Systems and Services (WONS)*, 2014, pp. 104–107.

[116] M. Ossmann, "Project ubertooth," *Retrieved*, vol. 18, p. 23, 2012.

[117] D. Spill and A. Bittau, "BlueSniff: Eve Meets Alice and Bluetooth.," *WOOT*, vol. 7, pp. 1–10, 2007.

[118] Y. Shaked and A. Wool, "Cracking the bluetooth pin," in *Proceedings of the 3rd international conference on Mobile systems, applications, and services*, 2005, pp. 39–50.

[119] Y. Lu and S. Vaudenay, "Faster correlation attack on Bluetooth keystream generator E0," in *Annual International Cryptology Conference*, 2004, pp. 407–425.

[120] Y. Lu, W. Meier, and S. Vaudenay, "The conditional correlation attack: A practical attack on bluetooth encryption," in *Annual International Cryptology Conference*, 2005, pp. 97–117.

[121] D. Chaum and H. Van Antwerpen, "Undeniable signatures," in *Conference on the Theory and Application of Cryptology*, 1989, pp. 212–216.

[122] J. Padgette, K. Scarfone, and L. Chen, *NIST special publication 800-121 re-vision 1, guide to bluetooth security*. .

[123] K. Lauter, "The advantages of elliptic curve cryptography for wireless security," *IEEE Wireless communications*, vol. 11, no. 1, pp. 62–67, 2004.

[124] C. Gehrmann, J. Persson, and B. Smeets, *Bluetooth Security*. Artech House, 2004.

[125] H. Martin, "Bluesnarfing@ CeBIT 2004, detecting and attacking bluetoothenabled cellphones at the Hannover Fairground," *Salzburg Research Forschungsgesellschaft mbH, Austria*, 2004.

[126] R. Bali, "Bluejacking Technology: Overview, Key Challenges and Initial Research."

[127] L. Owens, "Bluejacking: Bluetooth Graffiti," *Communications*, vol. 8, p. 9th, 2003.

[128] M. Tan and K. A. Masagca, "An investigation of Bluetooth security threats," in *Information Science and Applications (ICISA), 2011 International Conference on*, 2011, pp. 1–7.

[129] K. Munro, "Breaking into Bluetooth," *Network Security*, vol. 2008, no. 6, pp. 4–6, 2008.

[130] R. Bouhenguel, I. Mahgoub, and M. Ilyas, "Bluetooth security in wearable computing applications," in *High Capacity Optical Networks and Enabling Technologies, 2008. HONET 2008. International Symposium on*, 2008, pp. 182–186.

[131] K. M. Haataja and K. Hypponen, "Man-in-the-middle attacks on bluetooth: a comparative analysis, a novel attack, and countermeasures," in *Communications, Control and Signal Processing, 2008. ISCCSP 2008. 3rd International Symposium on*, 2008, pp. 1096–1102.

[132] B. P. Crow, I. Widjaja, J. G. Kim, and P. T. Sakai, "IEEE 802.11 wireless local area networks," *IEEE Communications magazine*, vol. 35, no. 9, pp. 116–126, 1997.

[133] E. D. Kaplan and C. Hegarty, Eds., *Understanding GPS: principles and applications*, 2nd ed. Boston: Artech House, 2006.

[134] "SafeTRIP - Satellite Applications For Emergency handling, Traffic Alerts, Road safety and Incident Prevention | indra." [Online]. Available: https://www.indracompany.com/en/indra/safetrip-satellite-applications-emergency-handling-traffic-alerts-road-safety-incident. [Accessed: 21-Jan-2019].

[135] S. Sasaki and T. Okazaki, "Feasibility study for telexistence on a ship - measurement of delay time of satellite communication," in *2016 IEEE International Conference on Systems, Man, and Cybernetics (SMC)*, 2016, pp. 001477–001482.

[136] B.-H. Kim, D.-G. Oh, Y.-J. Song, H. Kim, and H.-J. Lee, "Development of Ku-band Mobile Satellite Internet Access System," p. 7.

[137] U. H. Park, H. S. Noh, S. H. Son, K. H. Lee, and S. I. Jeon, "A Novel Mobile Antenna for Ku-Band Satellite Communications," *ETRI Journal*, vol. 27, no. 3, pp. 243–249, 2005.

[138] A. Roy-Chowdhury, J. S. Baras, M. Hadjitheodosiou, and S. Papademetriou, "Security issues in hybrid networks with a satellite component," *IEEE Wireless Communications*, vol. 12, no. 6, pp. 50–61, Dec. 2005.

[139] Y. Xiao, J. Liu, Y. Shen, X. Jiang, and N. Shiratori, "Secure Communication in Non-Geostationary Orbit Satellite Systems: A Physical Layer Security Perspective," *IEEE Access*, vol. 7, pp. 3371–3382, 2019.

[140] B. Harris and R. Hunt, "TCP/IP security threats and attack methods," *Computer communications*, vol. 22, no. 10, pp. 885–897, 1999.

[141] T. H. Jaeckle, "Determination of Integrity of Incoming Signals of Satellite Navigation System," US20150116147A1, 30-Apr-2015.

[142] R. Berry and A. Cook, "Detection of wireless data jamming and spoofing," US9466881B1, 11-Oct-2016.

[143] D. Moser, P. Leu, V. Lenders, A. Ranganathan, F. Ricciato, and S. Capkun, "Investigation of multi-device location spoofing attacks on air traffic control and possible countermeasures," in *Proceedings of the 22nd Annual International Conference on Mobile Computing and Networking*, 2016, pp. 375–386.

[144] A. H. Azhar, T. Tran, and D. O'Brien, "A gigabit/s indoor wireless transmission using MIMO-OFDM visible-light communications," *IEEE photonics technology letters*, vol. 25, no. 2, pp. 171–174, 2013.

[145] S. Rajagopal, R. D. Roberts, and S. Lim, "IEEE 802.15.7 visible light communication: modulation schemes and dimming support," *IEEE Communications Magazine*, vol. 50, no. 3, pp. 72–82, Mar. 2012.

[146] G. Pang, H. Liu, C.-H. Chan, and T. Kwan, "Vehicle location and navigation systems based on LEDs," *Proc. 5th World Congr. Intelligent Transport Systems*, pp. 12–16, 1998.

[147] G. C. PANG, "Dual use of LEDs: Signalling and communications in ITS," in *TOWARDS THE NEW HORIZON TOGETHER. PROCEEDINGS OF THE 5TH WORLD CONGRESS ON INTELLIGENT TRANSPORT SYSTEMS, HELD 12-16 OCTOBER 1998, SEOUL, KOREA. PAPER NO. 3035*, 1998.

[148] Z. MacHardy, A. Khan, K. Obana, and S. Iwashina, "V2X Access Technologies: Regulation, Research, and Remaining Challenges," *IEEE Communications Surveys Tutorials*, vol. 20, no. 3, pp. 1858–1877, thirdquarter 2018.

[149] A. Căilean and M. Dimian, "Current Challenges for Visible Light Communications Usage in Vehicle Applications: A Survey," *IEEE Communications Surveys Tutorials*, vol. 19, no. 4, pp. 2681–2703, Fourthquarter 2017.

[150] M. Foruhandeh, M. Uysal, I. Altunbas, T. Guven, and A. Gercek, "Optimal choice of transmission parameters for LDPC-coded CPM," in *2014 IEEE Military Communications Conference*, pp. 368–371, DOI 10.1109/MILCOM.2014.66, 2014.

[151] S. Kitano, S. Haruyama, and M. Nakagawa, "LED road illumination communications system," in *Vehicular*

Technology Conference, 2003. VTC 2003-Fall. 2003 IEEE 58th*, 2003, vol. 5, pp. 3346–3350.

[152] N. Kumar, "Smart and intelligent energy efficient public illumination system with ubiquitous communication for smart city," in *Smart Structures and Systems (ICSSS), 2013 IEEE International Conference on*, 2013, pp. 152–157.

[153] T. Saito, S. Haruyama, and M. Nakagawa, "A new tracking method using image sensor and photo diode for visible light road-to-vehicle communication," in *2008 10th International Conference on Advanced Communication Technology*, 2008, vol. 1, pp. 673–678.

[154] M. Y. Abualhoul, M. Marouf, O. Shagdar, and F. Nashashibi, "Platooning control using visible light communications: A feasibility study," in *Intelligent Transportation Systems-(ITSC), 2013 16th International IEEE Conference on*, 2013.

[155] N. Kumar, N. Lourenço, D. Terra, L. N. Alves, and R. L. Aguiar, "Visible light communications in intelligent transportation systems.," in *Intelligent Vehicles Symposium*, 2012, pp. 748–753.

[156] R. Roberts, P. Gopalakrishnan, and S. Rathi, "Visible light positioning: Automotive use case," in *Vehicular Networking Conference (VNC), 2010 IEEE*, 2010.

[157] Y. Ji, P. Yue, and Z. Cui, "VANET 2.0: integrating visible light with radio frequency communications for safety applications," in *International Conference on Cloud Computing and Security*, 2016, pp. 105–116.

[158] W. L. Junior, J. Costa, D. Rosário, E. Cerqueira, and L. A. Villas, "A Comparative Analysis of DSRC and VLC for Video Dissemination in Platoon of Vehicles."

[159] G. Blinowski, "Security issues in visible light communication systems," *IFAC-PapersOnLine*, vol. 48, no. 4, pp. 234–239, Jan. 2015.

[160] A. D. Wood, J. A. Stankovic, and S. H. Son, "JAM: A jammed-area mapping service for sensor networks," in *Real-Time Systems Symposium, 2003. RTSS 2003. 24th IEEE*, 2003, pp. 286–297.

[161] I. Marin-Garcia, A. M. Ramirez-Aguilera, V. Guerra, J. Rabadan, and R. Perez-Jimenez, "Data sniffing over an open VLC channel," in *Communication Systems, Networks and Digital Signal Processing (CSNDSP), 2016 10th International Symposium on*, 2016, pp. 1–6.

[162] D. Yucebas and H. Yuksel, "Power analysis based side-channel attack on visible light communication," *Physical Communication*, 2018.

[163] I. Marin-Garcia, V. Guerra, and R. Perez-Jimenez, "Study and Validation of Eavesdropping Scenarios over a Visible Light Communication Channel," *Sensors (Basel)*, vol. 17, no. 11, Nov. 2017.

[164] C. Rohner, S. Raza, D. Puccinelli, and T. Voigt, "Security in visible light communication: Novel challenges and opportunities," *Sensors & Transducers Journal*, vol. 192, no. 9, pp. 9–15, 2015.

[165] R. L. Rivest, "Chaffing and winnowing: Confidentiality without encryption," *CryptoBytes (RSA laboratories)*, vol. 4, no. 1, pp. 12–17, 1998.

[166] R. Wilson, D. Tse, and R. A. Scholtz, "Channel Identification: Secret Sharing Using Reciprocity in Ultrawideband Channels," *IEEE Transactions on Information Forensics and Security*, vol. 2, no. 3, pp. 364–375, Sep. 2007.

[167] M. Agiwal, A. Roy, and N. Saxena, "Next Generation 5G Wireless Networks: A Comprehensive Survey," *IEEE Communications Surveys Tutorials*, vol. 18, no. 3, pp. 1617–1655, thirdquarter 2016.

[168] G. Naik, J. Liu, and J.-M. J. Park, "Coexistence of wireless technologies in the 5 GHz bands: a survey of existing solutions and a roadmap for future research," *IEEE Communications Surveys & Tutorials*, vol. 20, no. 3, pp. 1777–1798, 2018.

[169] G. Mantas, N. Komninos, J. Rodriuez, E. Logota, and H. Marques, "Security for 5G communications," 2015.

[170] M. Foruhandeh and S. Aissa, "Efficiency of DC combination of rectified waveforms in energy harvesting systems," in *2015 IEEE International Conference on Ubiquitous Wireless Broadband (ICUWB)*, 2015, pp. 1–5.

[171] M. T. Garip, M. E. Gursoy, P. Reiher, and M. Gerla, "Congestion attacks to autonomous cars using vehicular botnets," in *NDSS Workshop on Security of Emerging Networking Technologies (SENT)*, 2015.

[172] D. Forsberg, H. Leping, K. Tsuyoshi, and S. Alanara, "Enhancing security and privacy in 3GPP E-UTRAN radio interface," in *Personal, Indoor and Mobile Radio Communications, 2007. PIMRC 2007. IEEE 18th International Symposium on*, 2007, pp. 1–5.

[173] P. Tyagi and D. Dembla, "Investigating the security threats in vehicular ad hoc networks (VANETs): towards security engineering for safer on-road transportation," in *Advances in Computing, Communications and Informatics (ICACCI, 2014 International Conference on*, 2014, pp. 2084–2090.

[174] M. Jeihani, S. NarooieNezhad, and K. B. Kelarestaghi, "Integration of a driving simulator and a traffic simulator case study: exploring drivers' behavior in response to variable message signs," *IATSS research*, vol. 41, no. 4, pp. 164–171, 2017.

[175] S. Banerjee, M. Jeihani, and D. Morris, "Impact of Work Zone Signage on Driver Speeding Behavior: A Driving Simulator Study (Ref.: Ms. No. 19-00714)." presented at the Transportation Research Board 98th Annual Meeting, no. 19-00714, 2019.

[176] Banerjee, S., Jeihani, M., Khadem, N. K., & Brown, D. D., "Units of information on dynamic message signs: a speed pattern analysis," *European Transport Research Review*, vol. 11, no. 1, p. 15, 2019.

[177] M. Jeihani and S. Banerjee, "Drivers' behavior analysis under reduced visibility conditions using a driving simulator," *Journal of Traffic and Logistics Engineering Vol*, vol. 6, no. 2, 2018.

[178] S. Banerjee, M. Jeihani, and R. Z. Moghaddam, "Impact of mobile work zone barriers on driving behavior on arterial roads," *Journal of Traffic and Logistics Engineering Vol*, vol. 6, no. 2, 2018.

[179] S. Dabiri, K. B. Kelarestaghi, and K. Heaslip, "Probe people and vehicle-based data sources application in the smart transportation." Preprint, DOI: 10.13140/RG.2.2.16487.7056.